\documentclass[journal]{IEEEtran}

\usepackage{ucs}
\usepackage[utf8x]{inputenc}
\usepackage[cmex10]{amsmath}
\usepackage{cite, amsfonts, amssymb, amsthm, bm, bbm, graphicx, relsize, multirow, booktabs, color,soul}
\usepackage[american]{babel}
\usepackage[T1]{fontenc}

\newtheorem{proposition}{Proposition}

\theoremstyle{definition}

\usepackage{algorithmic, algorithm}
\algsetup{indent=2em, linenodelimiter=.}

\theoremstyle{remark}

\DeclareMathOperator{\vect}{vec}

\DeclareMathOperator{\diag}{diag}
\DeclareMathOperator{\trace}{Tr}

\newcommand{\e}{\mathrm{e}}

\renewcommand{\j}{\mathrm{j}}

\newcommand*{\herm}{^{\mathsf{H}}}
\newcommand*{\transp}{^{\mathsf{T}}}

\title{MIMO OFDM Dual-Function Radar-Communication Under Error Rate and Beampattern Constraints}
\author{Jeremy~Johnston, Luca~Venturino,~\IEEEmembership{Senior~Member,~IEEE}, Emanuele~Grossi,~\IEEEmembership{Senior~Member,~IEEE}, Marco~Lops,~\IEEEmembership{Fellow,~IEEE}, Xiaodong~Wang,~\IEEEmembership{Fellow,~IEEE}   
\thanks{J.~Johnston and X.~Wang are  with the Department of Electrical Engineering, Columbia University, New York, NY 10027, United States (e-mail: j.johnston@columbia.edu; xw2008@columbia.edu).}
\thanks{L.~Venturino and E.~Grossi are with the Department of Electrical and Information Engineering, University of Cassino and Southern Lazio,  03043 Cassino, Italy, and with CNIT, 43124 Parma, Italy (e-mail: l.venturino@unicas.it; e.grossi@unicas.it).	}
\thanks{M.~Lops is with the Department of Electrical and Information Technology, University of Naples Federico II, 80138 Naples, Italy, and with CNIT, 43124 Parma, Italy (e-mail: lops@unina.it).}
\thanks{The work of L.~Venturino and E.~Grossi was supported by the research program ``Dipartimenti di Eccellenza 2018--2022'' sponsored by the Italian Ministry of Education, University, and Research (MIUR).}
}

\begin{document}
	\bstctlcite{BSTcontrol}		
	\maketitle	
	\IEEEpeerreviewmaketitle	
	\begin{abstract}
		In this work we consider a multiple-input multiple-output (MIMO) dual-function radar-communication (DFRC) system,  which senses multiple spatial directions and serves multiple users.  Upon resorting to an orthogonal frequency division multiplexing (OFDM) transmission format and a differential phase shift keying (DPSK) modulation, we study the design of the radiated waveforms and of the receive filters employed by the radar and the users. The approach is communication-centric, in the sense that a radar-oriented objective is optimized under constraints on the average transmit power, the power leakage towards specific directions, and the error rate of each user, thus safeguarding the communication quality of service (QoS).
		We adopt a unified design approach allowing a broad family of radar objectives, including both estimation- and detection-oriented merit functions. We devise a suboptimal solution based on alternating optimization of the involved variables, a convex restriction of the feasible search set, and minorization-maximization, offering a single algorithm for all of the radar merit functions in the considered family. Finally, the performance is inspected through numerical examples.
	\end{abstract}	
	
	\section{Introduction}
	The efficient use of the radio spectrum is a long-standing and challenging problem~\cite{Griffiths-2015,mazar2016radio}. Until recently, the frequency bands assigned to different wireless services have been kept mostly separate to avoid co-channel interference and hence simplify the system design; a static frequency planning, however, is inefficient. In the recent past, we have witnessed an increasing demand for mobile communication services that has driven the transition across three standards (3/4/5G) and fostered the proliferation of radar-based services in several areas (for example, industry automation, traffic monitoring, autonomous driving,  home surveillance, border patrolling, and earth monitoring); this has raised the cost for using any bandwidth slice and exacerbated the frequency shortage problem. Several solutions to improve the spectral efficiency have been implemented in communication networks,  including the use of sophisticated multiple access schemes and of  cognitive radios to allow a more dynamic spectrum management~\cite{Tarokh-2008,Proakis-book,Hanzo-2018}, the coordination of adjacent access points to enable a more aggressive spectrum reuse~\cite{Venturino2009,Venturino2011,Sanguinetti-2020}, and the exploitation of the spatial dimension for data encoding, modulation, and multiplexing~\cite{Venturino-2007,Fatema-2018,Albreem-2019}. Important technological advances have been also  made in the deployment of radar networks~\cite{Farina2020}, opening up the possibility of simultaneously scheduling multiple functions~\cite{Griffiths-2006} and implementing cognitive systems which sense the environment, learn relevant information, and then adapt to it~\cite{Griffiths-2019}; also, the use of multiple-input multiple-output (MIMO) digital transceivers~\cite{Li_Stoica_2009} and of the waverform diversity~\cite{Mokole-2016} have brought novel degrees of freedom for robust target detection~\cite{Venturino-MIMO-2011,Venturino-MIMO-2012}, adaptive signal processing~\cite{Stoica-2010,Li-2019}, and reconfigurable beam-pattern design~\cite{Rangaswamy-2019,Cui-2019}.	
	
	Cooperative spectrum sharing among licensed radar and communication systems is a key enabling technology for the efficient exploitation of the available bandwidth. The big divide among the solutions proposed so far is between radar and communication coexistence (RCC), wherein two distinct systems  negotiate their transmit/receive strategies to control the mutual interference, and dual-functional radar-communication (DFRC) systems, wherein  the radar and communication functions are combined in the same platform~\cite{Lops-2019,Liu-2020,Kim-2021}. RCC mainly results in a multi-objective optimization~\cite{Hongbin-2019,Liao-2019,Grossi-2020,Grossi-2021,Qian-TSP-2021} involving radar- and communication-oriented utilities with separable power constraints; widely-used performance measures are the data rate, the energy efficiency, and the error rate, at the communication side, and the signal-to-interference-plus-noise  ratio (SINR), the Cramér Rao bound on the variance of an unbiased estimator of a given unknown parameter, and the mutual information between the received signal and the target response, at the radar side.	A DFRC transceiver, conversely, can be implemented by complementing an existing communication module with a full-duplex receiver aimed at detecting the reflections from nearby scatterers~\cite{Grossi-TWC-2020,Grossi-TWC-2021}, in which case enabling the radar function may require the use of sophisticated receive strategies to cope with the imperfect ambiguity function of the communication signal. Alternatively, a message can be embedded into the waveforms radiated by an existing radar: effective strategies are the use of data-dependent coded pulses, the use of frequency/spatial index modulations, and the control of the sidelobes of the transmit beampattern towards the intended destinations~\cite{Hassanien-2016,Zhang-2018,Shlezinger-2020}. 	

	\subsection{Contribution of the Work}
    The joint design of the waveforms emitted by the DFRC transmitter and of the radar and communication receivers is a challenging and still debated  problem. The goal of this paper is to make a contribution in this domain; in particular, we consider an DFRC system employing an orthogonal frequency division multiplexing (OFDM) transmission format, wherein a MIMO transceiver simultaneously senses the environment and delivers a message to multiple users.  Due to its flexibility, OFDM is a good candidate technology to implement a DFRC system~\cite{Sturm-2011,Dingyou-2020}; indeed, OFDM is already widely-used in communications~\cite{Proakis-book}, while, more recently, has also received extensive attention in radar applications~\cite{Koivunen-2016}.	
	
	Previous works on OFDM-DFRC have mainly focused on single-antenna systems where the major degrees of freedom are the power allocation and the user scheduling among the available subcarriers~\cite{Hassanien-2019,Shi-2021} and/or the dynamical assignment of one function (either the radar or the communication) to each subcarrier~\cite{Koivunen-2019}. 
	The corresponding design strategies have consisted in maximizing the achievable communication sum-rate under a constraint on the radar mutual information \cite{Hassanien-2019}, minimizing the radiated power under constraints on both the radar and the communication mutual information \cite{Shi-2021}, or maximizing the sum of the radar and the communication mutual information \cite{Koivunen-2019}. 
	More recently, a massive MIMO OFDM system has been considered in~\cite{Temiz-2020,Temiz-2021,Temiz-2021-uplink}. In~\cite{Temiz-2020,Temiz-2021}, an access point simultaneously implements a short-range radar and serves multiple downlink users. Its antennas are separated into three groups, which radiate the radar waveforms, receive the echoes from the environment, and  radiate the data signals, respectively, and precoding strategies to enhance the system performance are studied. In~\cite{Temiz-2021-uplink}, instead, an access point simultaneously implements a short-range radar and receives signal from multiple uplink users, and the achievable performance are investigated under various operational conditions.

Differently from previous studies, in the present paper we focus  on the optimization of a radar-oriented  objective  function,  while safeguarding  the  communication  operation. In particular, the contribution can be summarized as follows.
\begin{itemize}
    \item Since the MIMO structure expands the number of degrees of freedom, both the transmitter and the receiver can be equipped with space-time filters that control the corresponding  beampatterns. Here we tackle the joint design of the DFRC transmitter and of the radar and user receivers and formulate a general resource allocation problem wherein the radar performance is optimized under constraints concerning the average transmit  power, the transmit beampattern (so as to limit the power leakage towards specific directions), and the error rate of each user (so as to safeguard the link quality).

    \item At the radar side, different directions can be inspected on each subcarrier, which allows handling multiple targets. We consider a broad family of merit functions for system design. This results in a unified design approach, which allows the system engineer  to reconfigure the radar task  at will and, also, to balance the radar performance on each subcarrier. For example, the considered family includes the \emph{quasi-arithmetic mean} of the radar SINRs on each subcarrier~\cite{Hardy_1934, Handbook-Means-2003, Lee-2017}, 
    the weighted-sum of the \emph{mutual information} between the received signal and the target response on each subcarrier~\cite{Blum-2007,Venturino-MIMO-2008,Nehorai-2010,Venturino-MIMO-2010},  the weighted-sum of the \emph{Fisher information} for the delay estimation on each subcarrier~\cite{Jajamovich_2010}, the weighted-sum of the \emph{detection probability} of the likelihood ratio-test on each subcarrier~\cite{Richards_2005}, and the weighted-sum of the two \emph{relative entropies} (also known as \emph{Kullback-Leibler divergences}) between the distributions of the received signal under the null hypothesis and its alternative on each subcarrier~\cite{EGrossi2012}.
    \item At the communication side, we do not assume full channel state information at the receiver, and a differential phase shift keying (DPSK) modulation is considered: this makes the transmit beampattern independent of the conveyed message and allows the users to employ an incoherent receiver for data demodulation. Needless to say, a coherent modulation scheme could be accounted for if channel state information were available. Also, we include in the model a different statistical characterization of the direct and indirect paths reaching each user. 
    \item Since the considered optimization is not convex, we derive an iterative algorithm---whose structure remains unaltered for all of the radar merit functions in the considered family---to compute a sub-optimal solution, which is based on the alternating optimization of the involved variables, a convex restriction of the feasible search set, and the minorization-maximization algorithm. The proposed procedure monotonically increases the objective function at each iteration and, hence, is convergent.
    \item Finally, we offer a set of curves showing some achievable radar and communication tradeoffs.
\end{itemize}  
    
    \subsection{Organization and Notation}
    The remainder of the paper is organized as follows. In Sec.~\ref{SEC:System-description},  the system description is presented. In Sec.~\ref{SEC:Problem-Formulation}, the proposed resource allocation problem is formulated and discussed, while a suboptimal solution is  derived in Sec.~\ref{Sec:Proposed algorithm}. In Sec.~\ref{SEC: Numerical_analysis}, some examples are given to illustrate the achievable tradeoffs between the radar and the communication operation. Concluding remarks are provided in Sec.~\ref{Sec:conclusions}. Finally, the Appendix contains the proofs of some of the presented results.
	
    In the following, $\mathbb R$, $\mathbb{R}_{+}$, and $\mathbb C$  are the set of real, non-negative and real, and complex numbers, respectively, while $\bar{\mathbb R}=\mathbb R \cup \{-\infty, \infty\}$ and $\bar{\mathbb R}_+={\mathbb R}_+ \cup \{\infty\}$. $\mathbb C^N$ and $\mathbb C^{N\times N}$ are the set of $ N\times 1$ vectors and $N\times N $ matrices with complex entries, respectively;$ (\,\cdot\,)^*$, $ (\,\cdot\,)\transp $, and $ (\,\cdot\,)\herm $ denote conjugate, transpose and conjugate transpose, respectively; $ \bm I_N $ is the $ N\times N $ identity matrix; $ \bm{1}_N $ and $ \bm{0}_N $ are the $N\times 1$ vectors with all-one and all-zero entries, respectively.  $ \trace\{\bm{X}\} $ is the trace of the square matrix $\bm{X}$; $\lambda_{\min}(\bm X)$ and $\lambda_{\max}(\bm X)$ are the minimum and maximum eigenvalue of the Hermitian matrix $\bm{X}$, respectively. $\vect\{\bm{X}\}$  is the vector obtained by stacking up the columns of $\bm{X}$. $\bm X \succeq 0$ and $\bm X \preceq 0$ means that $\bm X$ is Hermitian positive and negative semidefinite, respectively; if $\bm X_1$ and $\bm X_2$ are Hermitian matrices, then $\bm X_1\succeq \bm X_2$ means that $\bm X_1-\bm X_2\succeq0$.  $\diag( \{x_n\}) $ is the diagonal matrix with entries $x_1,\ldots,x_N$ on the main diagonal. We interchangeably use $f\left( x_{1},\ldots,x_{N}\right)$, $f\left(\bm{x}\right)$, and $f( \{x_{n}\})$ to denote a function $f$ of $\bm{x}=(x_{1},\ldots,x_{N})\transp$.  $f'$, $f''$, and $f'''$ are the first, second, and third derivative of $f$, respectively. $C^k$ denotes the differentiability class of order $k$. 
    $f^{-1}$, $\nabla_f$ and $\nabla_f^2$ are the inverse function, the gradient and the Hessian of $f$, respectively. $\mathbbm 1_\mathcal A$ is the indicator function of the condition $\mathcal A$, i.e., $\mathbbm 1_\mathcal A=1$, if $\mathcal A$ holds true, and $\mathbbm 1_\mathcal A=0$, otherwise in the notation paragraph. Finally, $\otimes$ and $\j$ indicate the Kronecker product and the imaginary unit, respectively.

	\section{System description}\label{SEC:System-description}	
	We consider an OFDM wireless system consisting of a DFRC transmitter, a co-located radar receiver, and $M$ communication users, each equipped with a linear array with closely-spaced antennas.\footnote{The following developments can be also extended to planar arrays.} We denote by  $N_{t}$, $N_{r}$, and $N_{m}$ the number of antennas at the transmitter, the radar receiver, and the $m$-th user, respectively. The OFDM symbol duration is much longer than the maximum propagation delay, so that a narrowband assumption holds  on each subcarrier~\cite{Sturm-2011,Koivunen-2016}. A subset of $K$ subcarriers is employed to simultaneously implement the radar and communication functions, while the other ones are not considered in this work. On each shared subcarrier, the transmitter aims to illuminate the direction of a prospective target while broadcasting a message to the users. We resort here to a DPSK modulation~\cite{Proakis-book} for data transmission; this is motivated by the facts that an incoherent receiver can be employed by each user and that the resulting transmit beampattern is independent of the selected data symbol.
	
	\subsection{Communication Side}
	\label{sec:commside}
	A direct  path and/or $Q_m\geq 0$ indirect paths (produced by as many far-field independent scatterers) can be present between the transmitter and the $m$-th user. Accordingly, its discrete-time received signal  on the $k$-th subcarrier  is modeled as 
	\begin{align}
		y_{k,m} &=  d_k \Big(\underbrace{\beta_{k,m,0}\trace\{\bm{W}_{k,m}\herm \bm{g}_{k,m}(\bar \phi_{m,0})\bm{s}_{k}\transp	( \phi_{m,0})	 \bm{U}_{k}\}}_{\text{direct link}}\notag \\
		&\quad  + \underbrace{\sum_{q = 1}^{Q_m} \beta_{k,m,q} \trace\{\bm{W}_{k,m}\herm 
			\bm{g}_{k,m}(\bar \phi_{m,q})\bm{s}_{k}\transp( \phi_{m,q})
			\bm{U}_{k}\}}_{\text{indirect links}}\Big)
		\notag \\
		&\quad +  \trace\{\bm{W}_{k,m}\herm \bm{Z}_{k,m}\}	 \label{eq: comSignalMatchFilter}
	\end{align}
	where: $\beta_{k,m,0}\in \mathbb{C}$ is the response of the direct path, while $\bar \phi_{m,0}$ and $\phi_{m,0}$ are the corresponding angles of arrival and departure, respectively;\footnote{Hereafter, all angles of arrival/departure are measured with respect to the array broadside direction and are positive when moving clockwise.} $\beta_{k,m,q}\in \mathbb{C}$, for $q=1,\ldots, Q_{m}$, is the response of the $q$-th indirect path, while $\bar \phi_{m,q}$ and $\phi_{m,q}$ are the corresponding angles of arrival and departure, respectively; $\bm{g}_{k,m}(\bar \varphi)\in \mathbb{C}^{N_m}$ and $ \bm{s}_{k}(\varphi) \in \mathbb{C}^{N_{t}}$ are the receive and transmit steering vectors, respectively, which are normalized to have  entries with unit magnitude; for example, if a uniform receive array is employed, we have  $	\bm{g}_{k,m}(\bar \varphi)=\big(1\;e^{-\j2\pi \frac{\mathsf{f}_{k}b_{m}}{c}\sin(\bar \varphi)}\;\cdots\; e^{-\j2\pi \frac{\mathsf{f}_{k}b_{m}}{c}\sin(\bar \varphi)(N_{m}-1)}\big)\transp$, where $\mathsf{f}_{k}$ is the center frequency of the $k$-th subcarrier, $b_{m}$ is the element spacing, and $c$ is the speed of light;	$\bm{U}_{k} \in{\mathbb C}^{N_{t}\times T}$ is the code matrix employed by the transmitter, which spans $T$ OFDM symbols;	$\bm W_{k,m} \in{\mathbb C}^{N_{m}\times T}$ is the filter employed by the user; 	$d_k\in\mathcal{D}=\{1,\e^{\j2\pi/D},\ldots,\e^{\j2\pi (D-1)/D}\}$ is the DPSK symbol to be broadcast, with $D$ being the cardinality of the constellation $\mathcal{D}$; and 	$\bm{Z}_{k,m}\in \mathbb{C}^{N_{m} \times T}$ is the disturbance vector.
	
	We assume that $\beta_{k,m,0}$  has a random phase, while its magnitude is deterministic and tied to the pathloss; $|\beta_{k,m,0}|= 0$ if no direct link is present, while $|\beta_{k,m,0}|> 0$ otherwise.   Also, we consider a Swerling I fluctuation model in each indirect path (as it includes the radar cross-section of the reflecting object), whereby $\beta_{k,m,q}$ is modeled as a circularly-symmetric Gaussian random variable with variance $ \sigma^2_{\beta,k,m,q}>0$~\cite{SkolnikBook}. Finally, we model the entries of $\bm{Z}_{k,m}$ as independent circularly-symmetric Gaussian random variables with variance $\sigma^2_{z,k,m}>0$.
	
	Upon defining   $\bm{u}_{k}=\vect\{\bm{U}_{k}\}$,  $\bm w_{k,m}=\vect\{\bm W_{k,m}\}$, $\bm z_{k,m}=\vect\{\bm Z_{k,m}\}$, and $\bm{G}_{k,m}(\bar \varphi,\varphi)=\bm{I}_{T}\otimes\bm{g}_{k,m}(\bar \varphi)\bm{s}_{k}\transp( \varphi)$, the signal in~\eqref{eq: comSignalMatchFilter} can be recast as
	\begin{equation}
		y_{k,m} =  d_k h_{k,m}+ \bm{w}_{k,m}\herm \bm{z}_{k,m} \label{eq: comSignalMatchFilter-2}
	\end{equation}
	where  $h_{k,m}=\sum_{q = 0}^{Q_m} \beta_{k,m,q} \bm{w}_{k,m}\herm \bm{G}_{k,m}(\bar \phi_{m,q},\phi_{m,q})  \bm{u}_{k}$
	is the channel response resulting from the superposition of the all paths reaching user $m$ on subcarrier $k$. Notice that $h_{k,m}$ is a complex random variable, and its magnitude follows a Rice distribution whose scale and shape parameters are~\cite{Proakis-book}
	\begin{subequations}\label{kappa_nu}
		\begin{align}	
			\nu_{k,m}&= |\beta_{k,m,0}|^2 \big| \bm{w}_{k,m}\herm \bm{G}_{k,m}(\bar \phi_{m,0},\phi_{m,0})\bm{u}_{k}\big|^2\notag\\
			&\quad +\sum_{q=1}^{Q_m} \sigma_{\beta,k,m,q}^2  \big| 
			\bm{w}_{k,m}\herm \bm{G}_{k,m}(\bar \phi_{m,q},\phi_{m,q}) \bm{u}_{k}
			\big|^2 \\
			\kappa_{k,m}&=  \frac{|\beta_{k,m,0}|^2 \big| \bm{w}_{k,m}\herm \bm{G}_{k,m}(\bar \phi_{m,0},\phi_{m,0})\bm{u}_{k}\big|^2}{\sum_{q=1}^{Q_m} \sigma_{\beta,k,m,q}^2  \big| 
				\bm{w}_{k,m}\herm \bm{G}_{k,m}(\bar \phi_{m,q},\phi_{m,q}) \bm{u}_{k}
				\big|^2}
		\end{align}
	\end{subequations}
	respectively. The parameter $\nu_{k,m}>0$ is the power received from all paths, while $\kappa_{k,m}\geq 0$ provides the ratio of the power along the direct path to that along the indirect paths. 
	
	Assuming that $h_{k,m}$ remains constant over two transmissions, we adopt an incoherent receiver to detect the phase offset over consecutive data symbols~\cite{Proakis-book}. We underline that such a receiver does not require the knowledge of $h_{k,m}$ for data demodulation. For $D=2$, the error probability for the $m$-th user on the $k$-th subcarrier is~\cite{Bargallo-1994}
	\begin{equation}\label{eq:E_D2}
		\text{E}_{k,m}=\frac{1+\kappa_{k,m}}{2(1+\kappa_{k,m}+\text{SNR}_{k,m})}\exp\left\{\frac{-\kappa_{k,m}\text{SNR}_{k,m}}{\kappa_{k,m}+\text{SNR}_{k,m}}\right\}
	\end{equation}
	where
	\begin{equation}\label{qe: SINR}
		\text{SNR}_{k,m}=\frac{\displaystyle  \nu_{k,m}}{\sigma^2_{z,k,m}\|\bm{w}_{k,m}\|^2}
	\end{equation}
	is the signal-to-noise-ratio (SNR). Notice that $\text{E}_{k,m}$ in~\eqref{eq:E_D2} is decreasing with both $\kappa_{k,m}$ and $\text{SNR}_{k,m}$ (see also~\cite[Fig.~1]{Bargallo-1994}). 
	For $D>2$, an integral expression of the  error probability $\text{E}_{k,m}$ is found in~\cite[Eq.~(5)]{Reed-1999} and omitted here for brevity; while this expression is more cumbersome, it still shows that $\text{E}_{k,m}$ is decreasing with both $\kappa_{k,m}$ and $\text{SNR}_{k,m}$. 
	
	\subsection{Radar Side}
	The radar inspects the direction $\psi_{k}$ on subcarrier $k$ and is aware of the presence of the self-interference (clutter) produced by  $J\geq0$ independent scatterers located in the directions $\theta_{1},\ldots,\theta_{J}$, with $\theta_{j}\neq \psi_{k}$ for any $j$ and $k$. Accordingly, the discrete-time signal received on the $k$-th subcarrier is modeled as 
	\begin{align}	
		&y_{k} =\underbrace{d_{k} \eta_{k} \trace\{\bm{W}_{k}\herm \bm{g}_{k}(\psi_{k})\bm{s}_{k}\transp(\psi_{k})
			\bm{U}_{k}\}}_{\text{target}} \notag 
		\\ &+ \underbrace{d_{k} \sum\limits_{j = 1}^J \alpha _{k,j} \trace\{\bm{W}_{k}\herm 
			\bm{g}_{k}(\theta_{j})\bm{s}_{k}\transp(\theta_{j})\bm{U}_{k}\}}_{\text{clutter}} + \trace\{\bm{W}_{k}\herm \bm{Z}_{k}\} \label{eq: subReceivedSignal}
	\end{align}
    where $\eta_{k} \in \mathbb{C}$ is the response of the target, $\alpha_{k,j} \in \mathbb{C}$ is the response of the $j$-th scatterer, $\bm{g}_{k}(\bar \varphi)\in \mathbb{C}^{N_{r}}$ is the receive steering vector,	$\bm{W}_{k} \in{\mathbb C}^{N_{r}\times T}$ is the filter employed by the radar receiver, and $\bm{Z}_{k}\in{\mathbb C}^{N_{r}\times T}$ is the disturbance vector.\footnote{Notice here that we use the same letter $y$ to denote the signal received by both the user $m$ and the radar. The understating is that the first subscript indexes the $k$-th subcarrier, while the second subscript, if present, identifies the $m$-th user, and, if absent, the radar receiver. A similar choice is made to denote the receive steering vector, the additive noise, and the receive filter.} 	We assume a Swerling I fluctuation for both the target and the clutter, whereby $\eta_{k}$ and $\alpha_{k,j}$ are independent circularly-symmetric Gaussian variables with variance  $\sigma_{\eta,k}^2>0$ and $\sigma_{\alpha,k,j}^2>0$, respectively~\cite{SkolnikBook}; accordingly, the unit-magnitude data symbol $d_{k}$ can be absorbed into  $\eta_{k}$ and $\alpha_{k,j}$ and does not play any role in the implementation of the radar receiver. Also, the entries of $\bm{Z}_{k}$ are modeled as independent circularly-symmetric Gaussian variables with variance $ \sigma^2_{z,k}>0$.
	
    Letting $\bm{w}_{k}=\vect\{\bm{W}_{k}\}$, $\bm z_{k}=\vect\{\bm Z_{k}\}$, and $\bm{G}_{k}(\theta_{j})=\bm{I}_{T}\otimes\bm{g}_{k}(\theta_{j})\bm{s}_{k}\transp( \theta_{j})$, the received signal can be rewritten as
	\begin{equation}	
		y_{k} =\eta_{k}\bm{w}_{k}\herm \bm{G}_{k}(\psi_{k})
		\bm{u}_{k}+ \sum\limits_{j = 1}^J \alpha _{k,j} \bm{w}_{k}\herm 
		\bm{G}_{k}(\theta_{j})\bm{u}_{k}+ \bm{w}_{k}\herm \bm{z}_{k}
	\end{equation}
	and the corresponding SINR is
	\begin{equation}\label{eq: case-1}
		{\rm SINR}_{k}=\frac{\sigma_{\eta,k}^2  \big|\bm{w}_{k}\herm\bm{G}_{k}(\psi_{k})\bm{u}_{k} \big|^2  
		}{ \sum_{j=1}^{J} \sigma_{\alpha,k,j}^2  \big| \bm{w}_{k}\herm\bm{G}_{k}(\theta_{j})\bm{u}_{k}\big|^2  +  \sigma^2_{z,k}\|\bm{w}_{k}\|^2}
	\end{equation}
	for $k=1,\ldots,K$. We now consider the following family of merit functions for system design
	\begin{equation}\label{radar-metric}
		f\left( {\rm SINR}_{1},\ldots,{\rm SINR}_{K}\right)
	\end{equation}
	where $f:\mathbb{R}_{+}^{K}\rightarrow \mathbb{R}_{+}$ is any increasing function that is either concave or minorized\footnote{The function $\zeta(\, \cdot\,| \bm x_0)$ minorizes $f$ at $\bm x_0$ if $f(\bm x)\geq \zeta(\bm x|\bm x_0)$, $\forall \bm x$, and $f(\bm x_0)= \zeta(\bm x|\bm x_0)$~\cite{Lange_2004}.} at any point $\bm x_0 \in\mathbb R_+^K$ by a concave function $\zeta(\,\cdot\,| \bm x_0)$. 
	Meaningful examples of radar merit functions will be provided later in Sec.~\ref{SEC_examples_radar_metric}.
    
    The following remarks are now in order. The radar pointing direction on each subcarrier can be that of a communication user or of another (prospective) object; likewise, the clutter may be caused by nearby users or other objects. The radar may inspect different directions on different subcarriers, thus handling  multiple targets. Interestingly enough, the above model and the following design methodology can be also extended to the case where the radar simultaneously inspects multiple directions on the same subcarrier employing as many receive filters: in this case, the family of merit functions in~\eqref{radar-metric} is modified to account for the individual SINR's on all inspected directions across all subcarriers; to keep the exposition concealed this generalization has been omitted.

	\subsection{Transmit Beampattern}\label{sec:TransmitBeampattern}
	The power radiated by the DFRC transmitter towards $\xi$ on subcarrier $k$ can be written as
	\begin{equation}
		\label{eq:txbp}
		\Delta_{k}\left(\bm{u}_{k}, \xi \right)=\frac{1}{T}\left\|\bm{s}_{k}\transp\left( \xi \right)\bm{U}_{k}\right\|^2\!\!=\frac{1}{T}\bm{u}_{k}\herm\bigr(\bm{I}_{T}\otimes \bm{s}_{k}^*\left( \xi \right)\bm{s}_{k}\transp\left( \xi \right)\bigl)\bm{u}_{k}.
	\end{equation}
	Notice that  $\Delta_{k}\left(\bm{u}_{k}, \xi \right)\leq N_{t} \mathcal{P}$, where $\mathcal{P}$ is the available power, with equality when all the power is assigned to subcarrier $k$ (i.e., $\bm{u}_p=\bm{0}_{N_{t}}$ for $p\neq k$ and $\|\bm{u}_{k}\|^2/T=\mathcal{P}$) and  $\bm{u}_{k}\propto \bm{1}_{T}\otimes \bm{s}_{k}^*(\xi)$.  It is desirable that the transmit beampattern in each subcarrier illuminate the directions corresponding to the prospective target and the connected users, while reducing the power leakage elsewhere, so as to limit the interference possibly caused to the radar receiver and to other co-channel systems operating nearby.  We denote by  $\xi_{k,1},\ldots,\xi_{k,L_{k}}$ the directions to be protected on subcarrier $k$, with $\xi_{k,\ell}\neq \psi_{k}$.
	
	\section{Problem Formulation}\label{SEC:Problem-Formulation}
	We assume here cognition of the surrounding environment, i.e., that the parameters $\{\bar \phi_{m,q},\phi_{m,q}\}$, $\{|\beta_{k,m,0}|\}$,  $\{\sigma_{\beta,k,m,q}^2\}$,  $\{\sigma^2_{z,k,m}\}$, $\{\theta_{j}\}$, $\{\sigma_{\alpha,k,j}^2 \}$, and $\{\sigma^2_{z,k}\}$ can be estimated~\cite{Griffiths-2019,Liu-2020}; on the other hand, the target powers $\{\sigma_{\eta,k}^2 \}$ may be set to a nominal value, as usual in radar design.
	
	For a given $\mathcal{P}$, we aim at maximizing a radar merit function of the form in~\eqref{radar-metric}, while guaranteeing a desired error rate for each user and constraining the transmit beampattern towards specific directions. The design variables are the transmit code $\{\bm{u}_{k}\}$, which allocate the power across the subcarriers and shape the transmit beampattern, and the receiver filters $\{\bm{w}_{k},\bm{w}_{k,m}\}$, which provide additional degrees of freedom for interference management. The problem to be solved is 
	\begin{equation} \label{eq: criterion1}
		\begin{aligned}
			\max_{\{\bm{u}_{k},\bm{w}_{k},\bm{w}_{k,m}\}}  & \; f\left( \big\{{\rm SINR}_{k}(\bm{u}_{k},\bm{w}_{k})\big\} \right) \\
			\text{s.t.}& \;\text{ C1:}\;  \frac{1}{T} \sum_{k=1}^{K}\|\bm{u}_{k}\|^2\leq \mathcal{P}\\	
			& \; \text{ C2:}\; \Delta_{k}\left(\bm{u}_{k}, \xi_{k,\ell} \right)\leq \delta_{k,\ell} N_{t} \mathcal{P},\; \forall\, k,\ell \\
			& \; \text{ C3:}\; \text{E}_{k,m}(\bm{u}_{k},\bm{w}_{k,m})\leq  \epsilon_{k,m},\quad \forall\, k,m
		\end{aligned}
	\end{equation}
	where $\delta_{k,\ell}\in[0,1]$ and $\epsilon_{k,m}\in(0,1/2)$. The above formulation can be readily modified to serve a different set of users on each subcarrier. For example, if user $m$ only needs to receive the message sent on the first subcarrier, then the  constraints on the error probability in the other subcarriers are simply removed; thus an orthogonal frequency division multiple access (OFDMA) can be obtained as a special case. 
	
	Problem~\eqref{eq: criterion1} is non-convex and hence difficult to solve. In the remaining part of this section we provide more insights into Problem~\eqref{eq: criterion1}; then in Section~\ref{Sec:Proposed algorithm} we propose a procedure to compute a suboptimal solution.  	
	
	\subsection{Examples of Radar Merit Functions}\label{SEC_examples_radar_metric}
	The family reported in~\eqref{radar-metric} encompasses several relevant merit functions. For example, we can consider the \emph{$p$-th power mean},  with $p\leq 1$, of the SINRs on each subcarrier~\cite{Hardy_1934, Handbook-Means-2003, Lee-2017}; in this case we have\footnote{We adopt the convention that $\frac{1}{0}=\infty$, $\frac{1}{\infty}=0$, and $\alpha+\infty=\infty$ for any $\alpha\geq0$; accordingly, for $p<0$, $f(\bm x)=0$ if $x_k=0$ for any $k\in\{1,\ldots,K\}$.}
	\begin{equation}\label{p-th-power-mean}
		f(\bm x) =\left(\sum_{k=1}^{K}\mu_{k}x_k^{p}\right)^{1/p}
	\end{equation}
	where $\{\mu_k\}$ are positive weights with  $\sum_{k=1}^{K}\mu_k=1$,  which  can be employed to give different priorities to different subcarriers.
	 The function in~\eqref{p-th-power-mean} is increasing and its concavity follows from the Minkowski's inequality~\cite{Handbook-Means-2003}[Ch.~4, Th.~9]. Also, its  value gets more biased towards its smallest argument as $p$ is decreased; in particular, it reduces to the arithmetic mean for $p=1$, the geometric mean for $p\rightarrow0$, the harmonic mean for $p=-1$, and $\min_{k\in\{1,\ldots,K\}}x_{k}$ for $p\rightarrow-\infty$.  
	
	We can also consider the \emph{quasi-arithmetic mean}  (also known as \emph{generalized mean}) of the SINRs generated by a continuous strictly monotone function $\gamma: \bar{\mathbb R}_+ \rightarrow \bar{\mathbb R}$~\cite{Hardy_1934, Handbook-Means-2003}, so that
	\begin{equation}
		f(\bm{x}) = \gamma^{-1}\left( \sum_{k=1}^{K}\mu_{k} \gamma(x_k) \right). \label{quasi-arithmetic_mean}
	\end{equation}
	This function is increasing and subsumes the $p$-th power mean for $\gamma(x)=x^p$ and the Geometric mean for $\gamma(x)=\ln x$; in the other cases, we can prove its concavity by exploiting the following proposition,\footnote{The result of this proposition still holds when the domain of $\gamma$ is a closed interval $[a,b]\subseteq \bar{\mathbb R}$, and the domain of $f$ is changed accordingly.} whose proof is provided in Appendix~\ref{proof_gen_mean}.
	\begin{proposition}
		Let $\gamma: \bar{\mathbb R}_+ \rightarrow \bar{\mathbb R}$ be a $C^4$ function  that is either strictly increasing and strictly concave or strictly decreasing and strictly convex. Then $f(\bm{x}) = \gamma^{-1}\bigl( \sum_{k=1}^{K}\mu_{k} \gamma(x_k) \bigr)$ is concave if and only if $\gamma'/\gamma''$ is convex. \label{proposition_gen_mean}
	\end{proposition}
	For example, $\gamma(x)=a^x$, with $a\in(0,1)$, satisfies the conditions of Proposition~\ref{proposition_gen_mean}, so that the resulting \emph{exponential mean}~\cite{Rennie_1991, Handbook-Means-2003} $f(\bm x) = \log_a\bigl( \sum_{k=1}^K \mu_k a^{x_k} \bigr)$  is concave. Also, $\gamma(x)=a^{1/x}$, with $a>1$, satisfies the conditions of Proposition~\ref{proposition_gen_mean}, so that the resulting \emph{radical mean}~\cite{Handbook-Means-2003}
	$f(\bm x) = \big(\log_a\big( \sum_{k=1}^K \mu_k a^{1/x_k}\big)\big)^{-1}$
	is concave.
	
	Furthermore, we can consider the weighted sum of the \emph{mutual information} between the received signal and the target response on each subcarrier, that is relevant in target classification. In this case, we have~\cite{Blum-2007,Venturino-MIMO-2008,Nehorai-2010,Venturino-MIMO-2010}
	\begin{equation}
		f(\bm x)=\sum_{k=1}^{K}\mu_{k}\ln\left(1+x_k\right)
	\end{equation}
	which is increasing and concave. 
	
	Additionally, we can consider the weighted-sum of the \emph{Fisher information} for the delay estimation on each subcarrier, which is related to the accuracy in target ranging. In this case, up to an irrelevant scaling factor, we have~\cite[cfr. Eq.~(35)]{Jajamovich_2010}
		\begin{equation}\label{eq-FI}
			f(\bm x)= \sum_{k=1}^K  \frac{\mu_k x_k^2}{1+x_k}
		\end{equation}	
		which is an increasing function. Moreover, since each term of the summation is convex and, therefore, lower-bounded by the tangent line, we have that $f$ is minorized at any $\bm x_0$ by the following concave (in fact, linear) function
		\begin{equation}
		    \zeta(\bm x|\bm x_0) = \sum_{k=1}^K \mu_k \left(\frac{x_{0,k}^2}{1+x_{0,k}} +\frac{2x_{0,k}+x_{0,k}^2}{(1+x_{0,k})^2} (x_k-x_{0,k})\right).
		\end{equation}	
	Moreover, we can consider the weighted sum of the detection probability of the likelihood ratio-test on each subcarrier. In this case, we have~\cite{Richards_2005}
	\begin{equation}\label{eq-Pd}
		f(\bm x)= \sum_{k=1}^K \mu_k P_{\text{fa},k}^{1/(1+x_k)}
	\end{equation}
	where $P_{\text{fa},k}$ is the probability of false alarm on the $k$-th subcarrier. This function is increasing, and, since
	\begin{subequations}
	 \begin{align}
	    \left(P_{\text{fa},k}^{1/(1+x_k)}\right)' &= - \frac{P_{\text{fa},k}^{1/(1+x_k)}\ln P_{\text{fa},k}}{(1+x_k)^2}\\
		\left(P_{\text{fa},k}^{1/(1+x_k)}\right)''  &= P_{\text{fa},k}^{1/(1+x_k)} \left(\frac{\left(\ln P_{\text{fa},k}\right)^2}{(1+x_k)^4} +\frac{2\ln P_{\text{fa},k}}{(1+x_k)^3}
			\right)\notag\\
			&\geq -\frac{(\sqrt{3}-3)^4 \e^{\sqrt{3}-3}}{\sqrt{3} \left(\ln P_{\text{fa},k}\right)^2}
		\end{align}%
	\end{subequations}
	a quadratic lower-bound for each term of the summation is readily obtained through Taylor's theorem. Therefore, $f$ is minorized at any $\bm x_0$ by the following concave function
		\begin{align}
		    \zeta(\bm x|\bm x_0) &= \sum_{k=1}^K \mu_k \Bigg(P_{\text{fa},k}^{1/(1+x_{0,k})} - \frac{P_{\text{fa},k}^{1/(1+x_{0,k})}\ln P_{\text{fa},k}}{(1+x_{0,k})^2}  \notag\\
		    &\quad \times (x_k-x_{0,k}) -\frac{(\sqrt{3}-3)^4 \e^{\sqrt{3}-3}}{2\sqrt{3} \left(\ln P_{\text{fa},k}\right)^2} (x_k-x_{0,k})^2\Bigg).
		\end{align}
	
	Finally, denote by $\mathcal{H}_{0}$ and $\mathcal{H}_{1}$ the null hypothesis (i.e., no target is present) and its alternative, respectively. We can consider the weighted-sum of the two \emph{relative entropies} between  $\mathcal{H}_{0}$ and $\mathcal{H}_{1}$ and between  $\mathcal{H}_{1}$ and $\mathcal{H}_{0}$  on each subcarrier, that can be used to control the average number of samples needed to make a decision in a sequential probability ratio test with given probabilities of detection and false alarm. In this case, we have~\cite[cfr. Sec.~III]{EGrossi2012}
	\begin{equation}\label{eq-KL}
		f(\bm x)= \sum_{k=1}^K \mu_k f_k(x_k)
	\end{equation}
	where $ f_k(x_k)=(1-2\omega_k) \ln(1+x_k) +x_k \frac{\omega_k x_k -(1-2\omega_k)}{1+x_k}$, with $\omega_k\in[0,1]$. This function is increasing, and, since
	\begin{subequations}
	 \begin{align}
	     f_k'(x_k) &= \frac{x_k(1+\omega_kx_k)}{(1+x_k)^2}\\
	     f_k''(x_k) & = \frac{1-(1-2\omega_k)x_k}{(1+x_k)^3} \geq - \frac{(1-2\omega_k)^3}{27(1-\omega_k)^2}
	 \end{align}%
	\end{subequations}
	each term of the summation is convex, if $\omega_k\geq 1/2$. Therefore, $f$ is minorized at any $\bm x_0$ by the following concave function
	\begin{align}
	    \zeta(\bm x|\bm x_0) &= \sum_{k=1}^K \mu_k \bigg(  f_k(x_{0,k}) + f_k' (x_{0,k}) (x_k-x_{0,k}) \notag\\
	    &\quad - \frac{(1-2\omega_k)^3}{54(1-\omega_k)^2} (x_k- x_{0,k})^2 \mathbbm{1}_{\{\omega_k<1/2\}}\bigg).
	\end{align}
	
	\subsection{Handling the Error Probability Constraint}	
	Varying $\bm{u}_{k}$ and/or $\bm{w}_{k,m}$ may have opposite effects on $\kappa_{k,m}$ and $\nu_{k,m}$ in~\eqref{kappa_nu}; accordingly,  the best tradeoff in terms of the error probability is in general not simple to assess. Interestingly, the dependency of $\text{E}_{k,m}$ upon $\bm{u}_{k}$ and $\bm{w}_{k,m}$ simplifies when $\kappa_{k,m}=\infty$ and $\kappa_{k,m}=0$, as discussed next.
	
	If $\kappa_{k,m}=\infty$, then only a direct path is present and no signal fading is observed; in this case, we have 
		\begin{equation}\label{E_inf_D2}
			\text{E}_{k,m}=\frac{1}{2}\e^{-\text{SNR}_{k,m}^{\rm d}}
		\end{equation}
		for $D=2$, and\footnote{DPSK loses about 3~dB with respect to the coherent PSK at large SNR's, and, in this regime, the nearest neighbor approximation to the error probability of the coherent PSK is tight.}
		\begin{equation}\label{E_inf_D>2}
			\text{E}_{k,m}\approx 2 Q\left(\sqrt{\text{SNR}_{k,m}^{\rm d} \sin^2\frac{\pi}{D}}\right)
		\end{equation}
		for $D>2$ and $\text{SNR}_{k,m}^{\rm d}\gg1$, where $Q(x)=\frac{1}{\sqrt{2\pi}}\int_x^\infty \e^{-t^2/2} dt$ and 
		\begin{equation}\label{eq: SINR-kappa-1}
			\text{SNR}_{k,m}^{\rm d}=\frac{|\beta_{k,m,0}|^2 \big| \bm{w}_{k,m}\herm \bm{G}_{k,m}(\bar \phi_{m,0},\phi_{m,0})\bm{u}_{k}\big|^2}{\sigma^2_{z,k,m}\|\bm{w}_{k,m}\|^2}.
		\end{equation}	
		
	If $\kappa_{k,m}=0$, then only the indirect paths are present and Rayleigh fading is observed; in this case, we have 
	\begin{equation}\label{E_0_D2}
			\text{E}_{k,m}=\frac{1}{2(1+\text{SNR}_{k,m}^{\rm i})}
		\end{equation}
		for $D=2$, and~\cite{Pauw_Schilling_1988}
		\begin{equation}\label{E_0_D>2}
			\text{E}_{k,m}\leq \frac{2\pi -\frac{2\pi}{D} +\sin\frac{2\pi}{D}}{2\pi \text{SNR}_{k,m}^{\rm i} \sin^2\frac{\pi}{D}}
		\end{equation}
		for $D>2$, where
		\begin{equation}\label{eq: SINR-kappa-0}
			\text{SNR}_{k,m}^{\rm i}=\frac{\displaystyle\sum_{q=1}^{Q_m} \sigma_{\beta,k,m,q}^2  \big| 
				\bm{w}_{k,m}\herm \bm{G}_{k,m}(\bar \phi_{m,q},\phi_{m,q}) \bm{u}_{k}
				\big|^2}{\sigma^2_{z,k,m}\|\bm{w}_{k,m}\|^2}.
	\end{equation}	
		
	For any $\bm{u}_{k}$ and  $\bm{w}_{k,m}$, $\kappa_{k,m}=\infty$ if the transmitter and user $m$ are in the line of sight and no close scatterers are present, while $\kappa_{k,m}=0$ if an obstacle blocks the direct path and nearby scatterers redirect the signal emitted by the transmitter towards user $m$. In all other cases, we can sub-optimally force $\kappa_{k,m}$ to be either $\infty$ or $0$ by operating on the receive filter $\bm{w}_{k,m}$, assuming that sufficient degrees of freedom are available. To be more specific, $\kappa_{k,m}$  can be set equal to infinity  by choosing $\bm{w}_{k,m}$ in the null space of the matrices $\{\bm{G}_{k,m}\herm(\bar \phi_{m,q},\phi_{m,q})\}_{q=1}^{Q_m}$, so as to zero-force the indirect signals: this is possible if the number of receive antennas is greater than the number of indirect paths (i.e., $N_{m}>Q_{m}$). Similarly, $\kappa_{k,m}$ can be set equal to zero by choosing $\bm{w}_{k,m}$  in the null space of $\bm{G}_{k,m}\herm(\bar \phi_{m,0},\phi_{m,0})$, so as to zero-force the direct signal: this is possible if the user has two or more antennas. Clearly, forcing $\kappa_{k,m}\in\{0,\infty\}$ amounts to adding a constraint into the optimization problem, that may lead to a sub-optimum solution.

	If $\kappa_{k,m}\in\{\infty,0\}$, it is verified from \eqref{E_inf_D2}--\eqref{eq: SINR-kappa-0} that upper bounding $\text{E}_{k,m}$ amounts to lower bounding 	\begin{equation}\label{SNR-def}
		\text{SNR}_{k,m}=\begin{cases}
			\text{SNR}_{k,m}^{\rm d},&\text{if } \kappa_{k,m}=\infty\\
			\text{SNR}_{k,m}^{\rm i} ,&\text{if } \kappa_{k,m}=0.
		\end{cases}
	\end{equation} Hence, the problem to be solved becomes 
	\begin{equation}\label{eq: criterion1RE}
		\begin{aligned}
			&\max_{\{\bm{u}_{k},\bm{w}_{k},\bm{w}_{k,m}\}}  \; f\left( \big\{{\rm SINR}_{k}(\bm{u}_{k},\bm{w}_{k})\big\} \right)\\
			\text{s.t.}& \;\text{ C1:}\;  \frac{1}{T} \sum_{k=1}^{K}\|\bm{u}_{k}\|^2\leq \mathcal{P}\\	
			& \; \text{ C2:}\; \Delta_{k}\left(\bm{u}_{k}, \xi_{k,\ell} \right)\leq \delta_{k,\ell} N_{t} \mathcal{P},\; \forall\, k,\ell\\
			&\; \text{ C3:}\;   {\text{SNR}}_{k,m}(\bm{u}_{k},\bm{w}_{k,m}) \geq \rho_{k,m},\; \forall \, m, k\\
			&\; \text{ C4:}\; \bm{w}_{k,m}\in\mathcal{W}_{k,m}
		\end{aligned}
	\end{equation}
	where  
	\begin{equation}
		\rho_{k,m}=\begin{cases}
			\rho_{k,m}^{\rm d},&\text{if } \kappa_{k,m}=\infty\\
			\rho_{k,m}^{\rm i},&\text{if } \kappa_{k,m}=0
		\end{cases}
	\end{equation}	
	is the minimum SNR required to satisfy the error rate constraint for the user $m$ on subcarrier $k$ and $
		\mathcal{W}_{k,m}=
		\big\{\bm{w}_{k,m}\in \mathbb{C}^{T N_{m}}:\,\kappa_{k,m}\in\{\infty,0\}\big\}$.
	
	Since the feasible search set of Problem~\eqref{eq: criterion1RE} is included in that of Problem~\eqref{eq: criterion1}, we have the following result.
	\begin{proposition}
		The solution to Problem~\eqref{eq: criterion1RE} provides a lower bound to the solution to Problem~\eqref{eq: criterion1}.
	\end{proposition} 
	
	\section{Proposed algorithm}\label{Sec:Proposed algorithm}	
	We compute here a suboptimal solution to~\eqref{eq: criterion1RE} by resorting to an alternating maximization. Starting from a feasible point, the objective function is maximized with respect to each of the block variables  $\{\bm{u}_{k}\}$, $\{\bm{w}_{k}\}$, and $\{\bm w_{k,m}\}$, taken in a cyclic order, while keeping the other ones fixed at their previous values. In the following, we discuss in detail the update of each block variable and the selection of the starting point. The overall procedure is summarized in Algorithm~\ref{proposed_alg} and is monotonically convergent, as the value of the objective function is not decreased at each iteration.  We underline here that convergence to a local/global optimum solution is not guaranteed as not all sub-problems are optimally solved.
	
	\begin{algorithm}[t]
		\caption{Proposed sub-optimal solution to Problem~\eqref{eq: criterion1RE}}
		\begin{algorithmic}[1]\label{proposed_alg}	
			\STATE Choose $\eta_\text{acc}>0$, $I_\text{max}>0$, and $\{\bm{u}_{k},\bm{w}_{k},\bm w_{k,m}\}$
			\STATE $i=0$ and $f^{(0)}= f\left( \big\{{\rm SINR}_{k}\big\} \right)$
			\REPEAT
			\STATE $i=i+1$
			\STATE Update $\{\bm{u}_{k}\}$ by solving~\eqref{eq: muvOptimaRE}		
			\STATE Update $\{\bm{w}_{k}\}$ as in~\eqref{opt-update-v}
			\STATE Update $\{\bm w_{k,m}\}$ as explained in Sec.~\ref{Sec: update-w}	
			\STATE $f^{(i)}= f\left( \big\{{\rm SINR}_{k}\big\} \right)$
			\UNTIL{$f^{(i)}-f^{(i-1)}<\eta_\text{acc} f^{(i)}$ or $i=I_\text{max}$}
		\end{algorithmic}
	\end{algorithm}
		
	\subsection{Update of the Transmit Code} 
	Upon defining ${\bm \varPsi}_{k,1}=\sigma^{2}_{\eta,k}\bm{G}_{k}\herm (\psi_{k})\bm{w}_{k} \bm{w}_{k}\herm \bm{G}_{k} (\psi_{k})$ and 
	${\bm \varPsi}_{k,2}=\sum_{j=1}^J\sigma^{2}_{\alpha,k,j}\bm{G}_{k}\herm (\theta_{j})\bm{w}_{k} \bm{w}_{k}\herm \bm{G}_{k} (\theta_{j})$, the problem to be solved is 
	\begin{equation}\label{eq: muvOptima}
			\max_{\{\bm{u}_{k}\}} \; 
			f\left( \left\{\frac{\bm{u}_{k}\herm{\bm \varPsi}_{k,1}\bm{u}_{k}}{\bm{u}_{k}\herm {\bm \varPsi}_{k,2}\bm{u}_{k}+\sigma^2_{z,k}\|\bm{w}_{k}\|^2}\right\} \right),\quad \text{s.t.} \;\text{ C1, C2, C3}
	\end{equation}
	Notice that the objective function in~\eqref{eq: muvOptima} is non-concave in the optimization variables, while the constraint C3 is non-convex.  
	
	To proceed, Problem~\eqref{eq: muvOptima} is first recast as
	\begin{equation}\label{eq: muvOptimaRe}
		\begin{aligned}
		\max_{\{\bm{u}_{k},x_k\}} & \; f(\bm x) \\
	    	\text{s.t.}& \;\text{ C1, C2, C3}\\
			&\; \text{ C5:}\;  \frac{\bm{u}_{k}\herm{\bm \varPsi}_{k,1}\bm{u}_{k}}{\bm{u}_{k}\herm {\bm \varPsi}_{k,2}\bm{u}_{k}+\sigma^2_{z,k}\|\bm{w}_{k}\|^2}\ge x_k, \; \forall\, k
		\end{aligned}
	\end{equation}
	where $x_1, \ldots,x_{K}$ are non-negative auxiliary variables and $\bm x =(x_1 \, \cdots \,x_{K})\transp$. Next, a convex restriction of C3 and C5 is derived. Let $\{\tilde{\bm{u}}_{k},\tilde{x}_{k}\}$ be the optimized variables at the previous iteration of Algorithm~\ref{proposed_alg}. Then, upon defining
	\begin{align}
		\bm{\varUpsilon}_{k,m}&=\frac{|\beta_{k,m,0}|^2}{\sigma^2_{z,k,m}\|\bm{w}_{k,m}\|^2} \notag \\ 	
		&\quad 	\times \bm{G}_{k,m}(\bar \phi_{m,0},\phi_{m,0})\bm{w}_{k,m} \bm{w}_{k,m}\herm\bm{G}_{k,m}\herm(\bar \phi_{m,0},\phi_{m,0})\notag \\ & \quad
		+\sum_{q=1}^{Q_m}  \frac{\sigma_{\beta,k,m,q}^2}{\sigma^2_{z,k,m}\|\bm{w}_{k,m}\|^2} \notag \\ 	&\quad 	\times \bm{G}_{k,m}(\bar \phi_{m,q},\phi_{m,q}) \bm{w}_{k,m}\bm{w}_{k,m}\herm
		\bm{G}_{k,m}\herm(\bar \phi_{m,q},\phi_{m,q})
	\end{align}
	we have~\cite{boyd2004convex} 
	\begin{align}
		\text{SNR}_{k,m}&\geq\tilde{\bm{u}}_{k}\herm \bm{\varUpsilon}_{k,m}	\tilde{\bm{u}}_{k}+2 \Re\left\{\tilde{\bm{u}}_{k}\herm\bm{\varUpsilon}_{k,m}\left(	\bm{u}_{k}-\tilde{\bm{u}}_{k}\right)\right\}\notag\\
		&=2 \Re\left\{\tilde{\bm{u}}_{k}\herm\bm{\varUpsilon}_{k,m}	\bm{u}_{k} \right\} - \tilde{\bm{u}}_{k}\herm \bm{\varUpsilon}_{k,m}	\tilde{\bm{u}}_{k}
	\end{align}
	where the inequality follows from the fact that $\text{SNR}_{k,m}=\bm{u}_{k}\herm \bm{\varUpsilon}_{k,m}	\bm{u}_{k} $ is a convex function of $\bm{u}_{k}$. At this point, we replace C3 with the following tighter constraint
	\begin{equation} \label{eq:C3R}
		\text{ C3R:}\; 2 \Re\left\{\tilde{\bm{u}}_{k}\herm\bm{\varUpsilon}_{k,m}	\bm{u}_{k} \right\} - \tilde{\bm{u}}_{k}\herm \bm{\varUpsilon}_{k,m}	\tilde{\bm{u}}_{k}\geq  \rho_{k,m}, \; \forall \, m, k.
	\end{equation}

	As to C5, first notice that it is active only when $x_k>0$, and, in this case, it can be rewritten as
	\begin{equation}
    	g_k(\bm{u}_{k},x_k) \geq \bm{u}_{k}\herm{\bm \varPsi}_{k,2}\bm{u}_{k}+\sigma^2_{z,k}\|\bm{w}_{k}\|^2 \label{C5-bis}
	\end{equation}
	where $g_k(\bm{u}_{k},x_k)  = \frac{1}{x_k}\bm{u}_{k}\herm{\bm \varPsi}_{k,1}\bm{u}_{k}$. We now have the following result, whose proof is provided in Appendix~\ref{proof_Prop-g1}.
	\begin{proposition}\label{Prop-g1} Let $\bm \varPsi \in \mathbb C^{N\times N}$, with $\bm \varPsi\succeq0$, and $g(\bm u, x)=\frac{1}{x}\bm u\herm\bm \varPsi \bm u$; then, $g: \mathbb C^{N} \times (0,\infty) \rightarrow \mathbb R$ is convex.
	\end{proposition}
	Then, exploiting the convexity of $g_k$, we have
	\begin{align}
		g_k\big(\bm{u}_{k},x_k\big)& \geq g_k\big(\tilde{\bm{u}}_{k},\tilde{x}_k\big)+ \Re \Bigg\{ \left(\frac{\partial g_k \big(\tilde{\bm{u}}_{k},\tilde{x}_k \big)}{\partial \bm{u}_{k}} \right)\herm \notag\\
		& \quad \times  \big(\bm{u}_{k}-\tilde{\bm{u}}_{k} \big) + \frac{\partial g_k (\tilde{\bm{u}}_{k},\tilde{x}_k )}{\partial x_k} \big( x_k-\tilde{x}_k\big) \Bigg\} \notag\\
		&= \frac{2}{\tilde{x}_k}\Re \big\{\tilde{\bm{u}}_{k}\herm{\bm \varPsi}_{k,1}\bm{u}_{k}\big\}-	\frac{x_k}{\tilde{x}_k^2}\tilde{\bm{u}}_{k}\herm {\bm \varPsi}_{k,1}\tilde{\bm{u}}_{k}\label{eq: psiFirstOrder}
	\end{align}
	if $\tilde x_k>0$, where the partial derivatives of $g_k$ are available from~\eqref{derivatives_g1_g1}. Therefore, from~\eqref{C5-bis} and~\eqref{eq: psiFirstOrder}, we can replace C5 with the following tighter constraint
	\begin{equation} 
		\text{C5R:}\;\begin{cases}
		\dfrac{2}{\tilde{x}_k}\Re \big\{\tilde{\bm{u}}_{k}\herm{\bm \varPsi}_{k,1}\bm{u}_{k}\big\}-	\dfrac{x_k}{\tilde{x}_k^2}\tilde{\bm{u}}_{k}\herm {\bm \varPsi}_{k,1}\tilde{\bm{u}}_{k} & \\
		\hfill \geq \bm{u}_{k}\herm{\bm \varPsi}_{k,2}\bm{u}_{k}+\sigma^2_{z,k}\|\bm{w}_{k}\|^2 , & \forall\, k: \tilde x_k>0\\
		x_k=0, & \forall\, k: \tilde x_k=0.
		\end{cases}\label{eq: C5Re}
    \end{equation}
    
	We now propose to solve the following restricted problem
	\begin{equation}\label{eq: muvOptimaRE}
			\max_{\{\bm{u}_{k},x_k\}} \; f(\bm x),\quad \text{s.t.} \;\text{ C1, C2, C3R, C5R}.
	\end{equation}
	
	If $f$ is concave, then \eqref{eq: muvOptimaRE} is a convex problem and can be solved  by using standard optimization techniques~\cite{boyd2004convex}; in this case, a solution to \eqref{eq: muvOptimaRE} is a feasible point for~\eqref{eq: muvOptima}; also, after updating $\{\bm{u}_{k},x_{k}\}$ as in~\eqref{eq: muvOptimaRE},  the value of the objective function in \eqref{eq: muvOptima} is not decreased.
	
	If $f$ is not concave but it can be minorized at any point by a concave function, we can sub-optimally solve Problem~\eqref{eq: muvOptimaRE} via a minorization-maximization algorithm~\cite{Lange_2004}. Specifically, starting from $\bm x^{(0)}= (\tilde x_{1},\ldots,\tilde x_{K})\transp$, a sequence of feasible points is generated by the following induction: given $\bm x^{(i-1)}$, choose $\bm x^{(i)}$ as the solution to the following convex problem
	\begin{equation}\label{eq: muvOptimaRE-MM}
			\max_{\{\bm{u}_{k},x_k\}} \; \zeta\bigl(\bm x | \bm x^{(i-1)}\bigr),\quad			\text{s.t.} \;\text{ C1, C2, C3R, C5R}.
	\end{equation}
	The solution to \eqref{eq: muvOptimaRE-MM} is a feasible point for~\eqref{eq: muvOptima}; also, after updating $\{\bm{u}_{k},x_{k}\}$ as in~\eqref{eq: muvOptimaRE-MM},  we have
	\begin{equation}
		f\bigl(\bm x^{(i)}\bigr) \geq \zeta\bigl(\bm x^{(i)} | \bm x^{(i-1)}\bigr) \geq \zeta \bigl(\bm x^{(i-1)} | \bm x^{(i-1)}\bigr) = f\bigl( \bm x^{(i-1)}\bigr)
	\end{equation}
	whereby $\{ f(\bm x^{(i)})\}_{i\in\mathbb N}$ is a non decreasing sequence.  Since solving~\eqref{eq: muvOptimaRE} is part of an  alternating-maximization algorithm, it is not necessary to iterate the maximization of $g(\bm x | \bm x^{(i-1)})$ in~\eqref{eq: muvOptimaRE-MM} until convergence and we can just proceed to update the other block variables $\{\bm{w}_{k}\}$ and $\{\bm w_{k,m}\}$ after only one or few steps of inner minorization-maximization.	The assumption that  $f(\cdot)$  be (locally) minorized by a concave function is quite mild in practice, and, indeed, Sec.~III-A lists several meaningful merit functions possessing such property. Clearly, a possible difficulty may come from finding the appropriate surrogate function that minorizes the desired objective function. Indeed, a good surrogate function should try to closely follow the shape of the objective function so as to yield a faster convergence rate in the minorization-maximization algorithm; on the other hand, it should be also possess a simple structure to reduce the computational cost per iteration.

	\subsection{Update of the Radar Receive Filters}\label{SEC_optimal_v}
	The filter $\bm{w}_{k}$ only comes into play in the objective function. Since $f$ is increasing, the optimal $\bm{w}_{k}$ must maximize ${\rm SINR}_{k}$. This problem is separable for each user and subcarrier and admits a closed form solution. Indeed, we have
	\begin{align}\label{eq: criterion1v}
		{\rm SINR}_{k}&=	\frac{\sigma_{\eta,k}^2 \bm{w}_{k}\herm  \bm{G}_{k}(\psi_{k}) \bm{u}_{k}\bm{u}_{k}\herm \bm{G}_{k}(\psi_{k}) \bm{w}_{k}}{\bm{w}_{k}\herm \bm{\Phi}_{k}(\bm{u}_{k}) \bm{w}_{k}}\notag \\
		&\leq \sigma_{\eta,k}^2\bm{u}_{k}\herm \bm{G}_{k}\herm(\psi_{k}) \bm{\Phi}_{k}^{-1}(\bm{u}_{k}) \bm{G}_{k}(\psi_{k})\bm{u}_{k}
	\end{align}	
	where $
		\bm{\Phi}_{k}(\bm{u}_{k})=\sum_{j=1}^{J} \sigma_{\alpha,k,j}^2 \bm{G}_{k}(\theta_{j}) \bm{u}_{k}\bm{u}_{k}\herm\bm{G}_{k}(\theta_{j})+\sigma^2_{z,k}\bm I_{N_{r}}$,	and the upper bound is achieved (up to an irrelevant scaling factor)  when~\cite{Gershman_2006} 
	\begin{equation}\label{opt-update-v}
		\bm{w}_{k}=\bm{\Phi}_{k}^{-1}(\bm{u}_{k}) \bm{G}_{k}(\psi_{k})\bm{u}_{k}.
	\end{equation}	
	
	\subsection{Update of the User Receive Filters} \label{Sec: update-w}
	Let $\bm{\Pi}_{k,m}^{\rm d}$ be equal to the projector onto the orthogonal  complement of the subspace spanned by the vectors $\{\bm{g}_{k,m}(\bar \phi_{m,q})\}_{q=1}^{Q_m}$, if $Q_{m}>0$, and to $\bm{I}_{N_{m}}$ otherwise; also, let $\bm{\Pi}_{k,m}^{\rm i}$ be equal to the projector onto the orthogonal complement of the subspace spanned by  $\bm{g}_{k,m}(\bar \phi_{m,0})$, if $|\beta_{k,m,0}|>0$, and to $\bm{I}_{N_{m}}$ otherwise; finally, let 
	\begin{align}\label{Xi-def}
		\bm{\Xi}_{k,m}=	\dfrac{|\beta_{k,m,0}|^2}{\sigma^2_{z,k,m}}\bm{G}_{k,m}(\bar \phi_{m,0},\phi_{m,0})\bm{u}_{k}\bm{u}_{k}\herm\bm{G}_{k,m}\herm(\bar \phi_{m,0},\phi_{m,0})\notag \\ +\sum_{q=1}^{Q_m} \dfrac{\sigma_{\beta,k,m,q}^2}{\sigma^2_{z,k,m}}  \bm{G}_{k,m}(\bar \phi_{m,q},\phi_{m,q}) \bm{u}_{k}  \bm{u}_{k}\herm
		\bm{G}_{k,m}\herm(\bar \phi_{m,q},\phi_{m,q}).
	\end{align}
	Then, we have the following result, whose proof is reported in Appendix~\ref{proof_Prop-w}.
	\begin{proposition}\label{Prop-w}
		If Problem~\eqref{eq: criterion1RE}  is feasible, then the optimal $\bm{w}_{k,m}$ is proportional to  the eigenvector  corresponding to the largest eigenvalue of $\bm{\Xi}_{k,m}^{\rm d}=\big(\bm{I}_{T}\otimes\bm{\Pi}_{k,m}^{\rm d}\big)\bm{\Xi}_{k,m}\big(\bm{I}_{T}\otimes\bm{\Pi}_{k,m}^{\rm d}\big)$, if $\kappa_{k,m}=\infty$, and of	$\bm{\Xi}_{k,m}^{\rm i}=\big(\bm{I}_{T}\otimes\bm{\Pi}_{k,m}^{\rm i}\big)\bm{\Xi}_{k,m}\big(\bm{I}_{T}\otimes\bm{\Pi}_{k,m}^{\rm i}\big)$,
		if $\kappa_{k,m}=0$, for $m=1,\ldots,M$ and $k=1,\ldots,K$.
	\end{proposition}
	According to Proposition~\ref{Prop-w}, we compute here two candidate solutions for $\bm{w}_{k,m}$; one solution is proportional to the eigenvector corresponding to the maximum eigenvalue of $\bm{\Xi}_{k,m}^{\rm d}$	(so that $\kappa_{k,m}=\infty$), while the other to the eigenvector corresponding to the maximum eigenvalue of $\bm{\Xi}_{k,m}^{\rm i}$ (so that $\kappa_{k,m}=0$). When both solutions are feasible, the one providing the lower error probability is selected; otherwise, we simply keep the one which is feasible.
	
	Notice in passing that, by leveraging Proposition~\ref{Prop-w}, we can also obtain the following side result on the feasibility of C3, whose proof is reported in Appendix~\ref{proof_Prop-C3}.
	\begin{proposition}\label{Prop-C3}
	A necessary condition for the feasibility of C3 in Problem~\eqref{eq: criterion1RE} is
    $\|\bm{U}_{k}\herm\bm{s}_{k}^{*}(\phi_{m,0})\|\neq 0$,  if $\kappa_{k,m}=\infty$, and $\max_{q\in\{1,\ldots,Q_{m}\}}\|\bm{U}_{k}\herm \bm{s}_{k}^{*}( \phi_{m,q})\|>0$, if $\kappa_{k,m}=0$, for $m=1,\ldots,M$ and $k=1,\ldots,K$. If $\text{Rank}\{\bm{U}_{k}\}=N_{t}$, a sufficient condition for the feasibility of C3 in Problem~\eqref{eq: criterion1RE} is 
		\begin{align}\label{C3_feasibility}
			&\lambda_{\min}(\bm{U}_{k}\bm{U}_{k}\herm)\geq \notag\\
			&\begin{cases}\displaystyle \frac{\rho_{k,m}^{\rm d} \sigma^2_{z,k,m}/N_{t}}{\displaystyle |\beta_{k,m,0}|^2 \bm{g}\herm(\bar \phi_{m,0})\bm{\Pi}_{k,m}^{\rm d}\bm{g}_{k,m}(\bar \phi_{m,0})}, & \text{if } \kappa_{k,m}=\infty\\[10pt]
				\displaystyle	\frac{\rho_{k,m}^{\rm i} \sigma^2_{z,k,m}/N_{t}}{\displaystyle\max_{q}|\beta_{k,m,q}|^2  \bm{g}\herm(\bar \phi_{m,q})\bm{\Pi}_{k,m}^{\rm i}\bm{g}_{k,m}(\bar \phi_{m,q})}, & \text{if } \kappa_{k,m}=0
			\end{cases}
		\end{align}
		for $m=1,\ldots,M$ and $k=1,\ldots,K$.
	\end{proposition}	
	
\subsection{Selection of the Starting Point}\label{SEC:start-point}
Next we outline a possible method to obtain a starting point for Algorithm~\ref{proposed_alg}. To this end, consider the following problem
	\begin{equation} \label{feasprob}
		\begin{aligned}
		\max_{t,\{\bm{u}_{k}\},\{\bm{w}_{k,m}\}} & \; t\\
		\text{s.t.} & \; \text{C1, C2, C4}\\
			& \; \overline{\text{C3}}\text{:} \; {\text{SNR}}_{k,m}(\bm{u}_{k},\bm{w}_{k,m})/\rho_{k,m} \geq t.
		\end{aligned}
	\end{equation}
If the optimal $t$ is not lower than $1$, then the corresponding variables $\{\bm{u}_{k}\}$ and $\{\bm{w}_{k,m}\}$ together with the radar receive filters $\{\bm{w}_{k}\}$ obtained as by-product from~\eqref{opt-update-v} are a feasible point for Algorithm~\ref{proposed_alg}. Since~\eqref{feasprob} is an NP-hard program~\cite{1634819}, we resort here to an alternating maximization of the block variables $\{t,\,\bm{u}_{k}\}$ and $\{\bm{w}_{k,m}\}$. Given $\{\bm{w}_{k,m}\}$, we can update $\{t,\,\bm{u}_{k}\}$ by solving the following problem
\begin{equation} \label{feasprob_no_w}
	\begin{aligned}
	\max_{t,\{\bm{u}_{k}\}} & \; t, \quad \text{s.t.}\; \text{C1},\, \text{C2},\,\overline{\text{C3}},
	\end{aligned}
\end{equation}
which can be tackled, similarly to what was done for Problem~\eqref{eq: muvOptimaRe}, by introducing a convex restriction of $\overline{\text{C3}}$.  Also, given $\{t,\,\bm{u}_{k}\}$,  we can update $\{\bm{w}_{k,m}\}$ as described in Sec.~\ref{Sec: update-w}. At the beginning, $\{\bm{u}_{k}\}$ can be randomly selected and normalized to meet C1. Also, $\{\bm{w}_{k,m}\}$ can be initialized as follows. First we randomly decide whether user $k$ will utilize the direct path or the indirect paths (if both present) on the  $k$-th subcarrier; if the direct path is used, then $\bm{w}_{k,m}= (\bm{I}_{T}\otimes\bm{\Pi}_{k,m}^{\rm i})( \bm{I}_{T}\otimes\bm{g}_{k,m}(\bar \phi_{m,0}))$, otherwise, $\bm{w}_{k,m}= (\bm{I}_{T}\otimes\bm{\Pi}_{k,m}^{\rm d})( \bm{I}_{T}\otimes\bm{g}_{k,m}(\bar \phi_{m,\hat{q}}))$, where $\hat{q}=\max_{q\in\{1,\ldots,Q_m\}}\sigma^2_{\beta,k,m,q}$.

    \section{Numerical analysis}\label{SEC: Numerical_analysis} 
     
     We consider an OFDM system using a power $\mathcal P=20$~dBW on the shared subcarriers.  The  center frequency of the $k$-th subcarriers is $\mathsf{f}_{k} = \mathsf{f}_{0} + (k-1)\Delta \mathsf f$, where  $\mathsf f_0=2$~GHz and $\Delta \mathsf f=100$~KHz, while each array has a uniform element spacing of $c/(2\max_k{\mathsf{f}_{k}})$.  At the communication side, we set $\sigma_{z,k,m}^2=-150$~dBW and $\epsilon_{k,m}=\epsilon$. The users have either one direct path or two indirect paths (they will be referred to as the direct and indirect users, respectively) with the corresponding angles of arrival/departure sampled from the uniform distribution on $[-\pi/3,\pi/3]$. The response of the direct path is set to  have $|\beta_{k,m,0}|^2=-130$~dB, while that of the indirect paths is chosen to have $\sum_{q=1}^{Q_{m}}\sigma_{\beta,k,m,q}^2=-130$~dBW. Unless otherwise stated, we consider two direct and two indirect users. At the radar side,  a different direction randomly chosen in $[-\pi/3,\pi/3]$ is inspected on each subcarrier. Also, we set  $\sigma_{z,k}^2=-150$~dBW, $\sigma_{\eta,k}^2=-160$~dB, and $J=4$, while the strength of each clutter element is assumed to be equal and adjusted according to a given signal-to-clutter ratio (SCR), defined as\footnote{The parameter $\text{SCR}$ is the ratio between the power of the target response and the sum-power of the response of each clutter element; accordingly, it does not account for the combined effects of the transmit and receive filters.} $\text{SCR} = \sigma_{\eta,k}^2 / \sum_{j=1}^{J} \sigma_{\alpha,k,j}^2$. As to the transmit beampattern, we set $\delta_{k,\ell}=\delta$ and $L_{k}=L$, while the protected directions are randomly chosen $[-\pi/3,\pi/3]$. Unless otherwise stated, we consider $L=6$. The other parameters are given in Table~\ref{tab:simparams}.
     Algorithm~\ref{proposed_alg} is implemented with $\eta_\text{acc} = 10^{-4}$ and $I_\text{max}=2000$, while the block variables $\{\bm u_k\}$, $\{\bm w_k\}$, and $\{\bm w_{k,m}\}$ are initialized as in Sec.~\ref{SEC:start-point}; finally, the auxiliary variable ${\tilde{x}}_k$ in Problem~\eqref{eq: muvOptimaRE} is initialize to $\mathrm{SINR}_{k}$.
     
     \begin{table}[t]
		\caption{}
		\label{tab:simparams}
		\centering
		\begin{tabular}{lll}  
			\toprule
			Parameter&Value&Description\\
			\midrule
			$N_t$ & 11& number of transmit antennas\\ 
			$N_r$ & 4 & number of radar receive antennas\\   
			$N_c$ & 4 & number of user receive antennas\\   
			$K$ & 4 & number of subcarriers \\   
			$T$ & 2 & number of time slots\\  
			$D$ & 2 & constellation size \\
			\bottomrule
		\end{tabular}
	\end{table}
	
  \subsection{Examples}
\begin{figure}[t]
		\centering
		\centerline{\includegraphics[width=0.9\columnwidth]{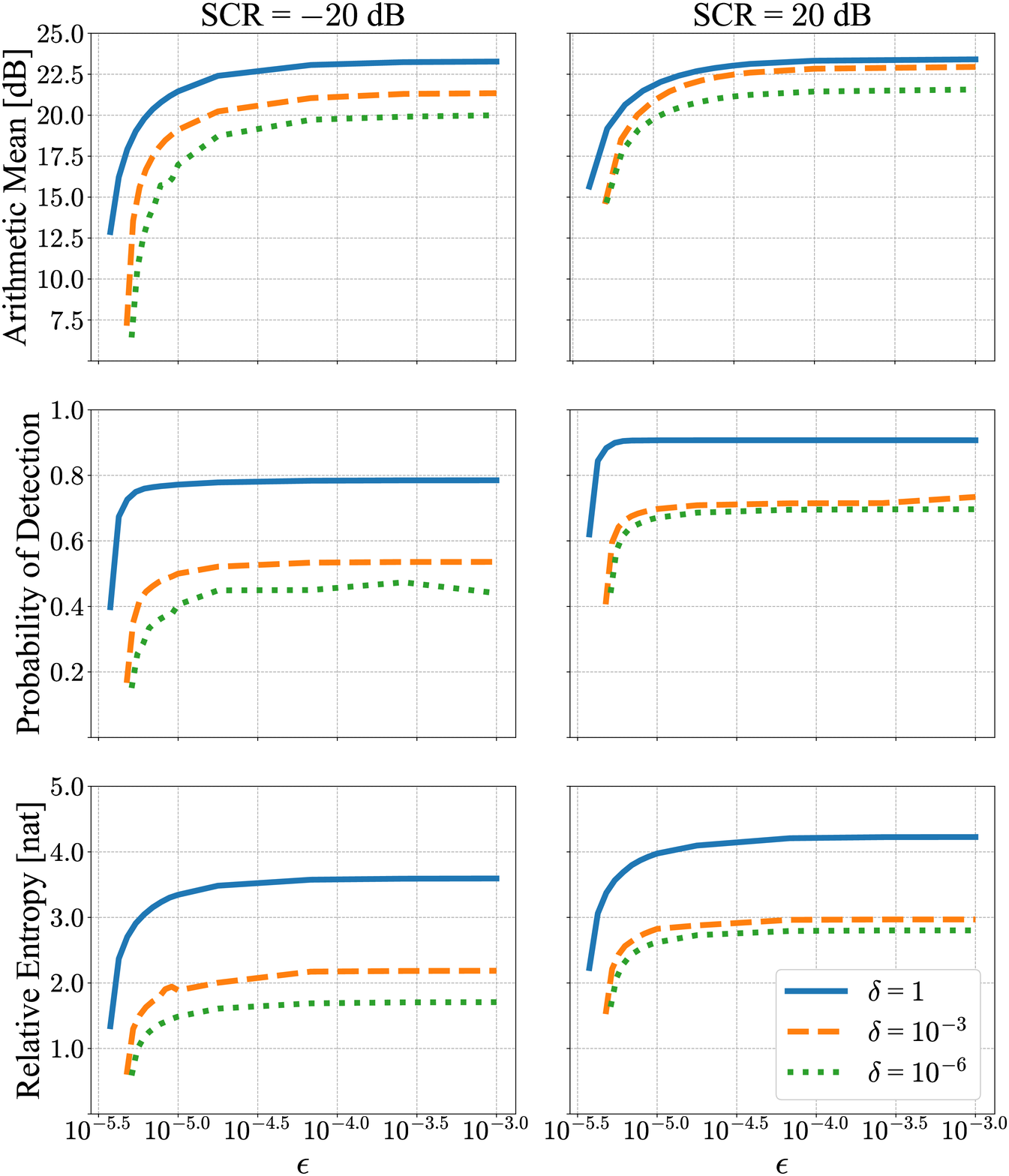}}
		\caption{Arithmetic mean  of the radar SINRs (top), probability of detection with $P_{\mathrm{fa},k}=10^{-4}$ (middle), and the relative entropy with $\omega_k=0$ (bottom) vs. the constraint~$\epsilon$ on the user error rate for $\delta=10^{-6}, 10^{-3}, 1$ when SCR of $-20$~dB (left) and $20$~dB (right).}
		\label{fig:SINR}
	\end{figure}

First we examine the interplay between the radar and communication operations, which compete here for the same physical resources.  We consider three radar merit functions: the arithmetic mean, i.e., the merit function in~\eqref{p-th-power-mean} with $p=1$; the detection probability in~\eqref{eq-Pd} with $P_{\text{fa},k}=10^{-4}$; and the relative entropy in~\eqref{eq-KL} with $\omega_{k}=0$. In all cases, we set $\mu_k = 1/K$.  Fig.~\ref{fig:SINR} reports these radar merit functions versus the constraint~$\epsilon$ on the user error rate for $\delta=10^{-6}, 10^{-3}, 1$; two SCRs are considered, namely, $-20$~dB and $20$~dB. The curves are obtained by averaging over $10$ problem instances.	A lower objective is attained when $\delta$  and/or SCR decrease, since the system consumes more degrees of freedom to reduce the power leakage towards the clutter and the protected directions; notice here that $\delta = 1$ is tantamount to removing  C2. There is an evident trade-off between the radar and communication performance; in particular, the radar performance sharply drops when $\epsilon$ gets lower than $10^{-5}$, as most of the system resources are employed for data transmission; instead, the communication function only marginally restrains the radar performance if $\epsilon$ is larger than $10^{-3.5}$.

	 \begin{figure}[t]
		\centering
		\centerline{\includegraphics[width=1\columnwidth]{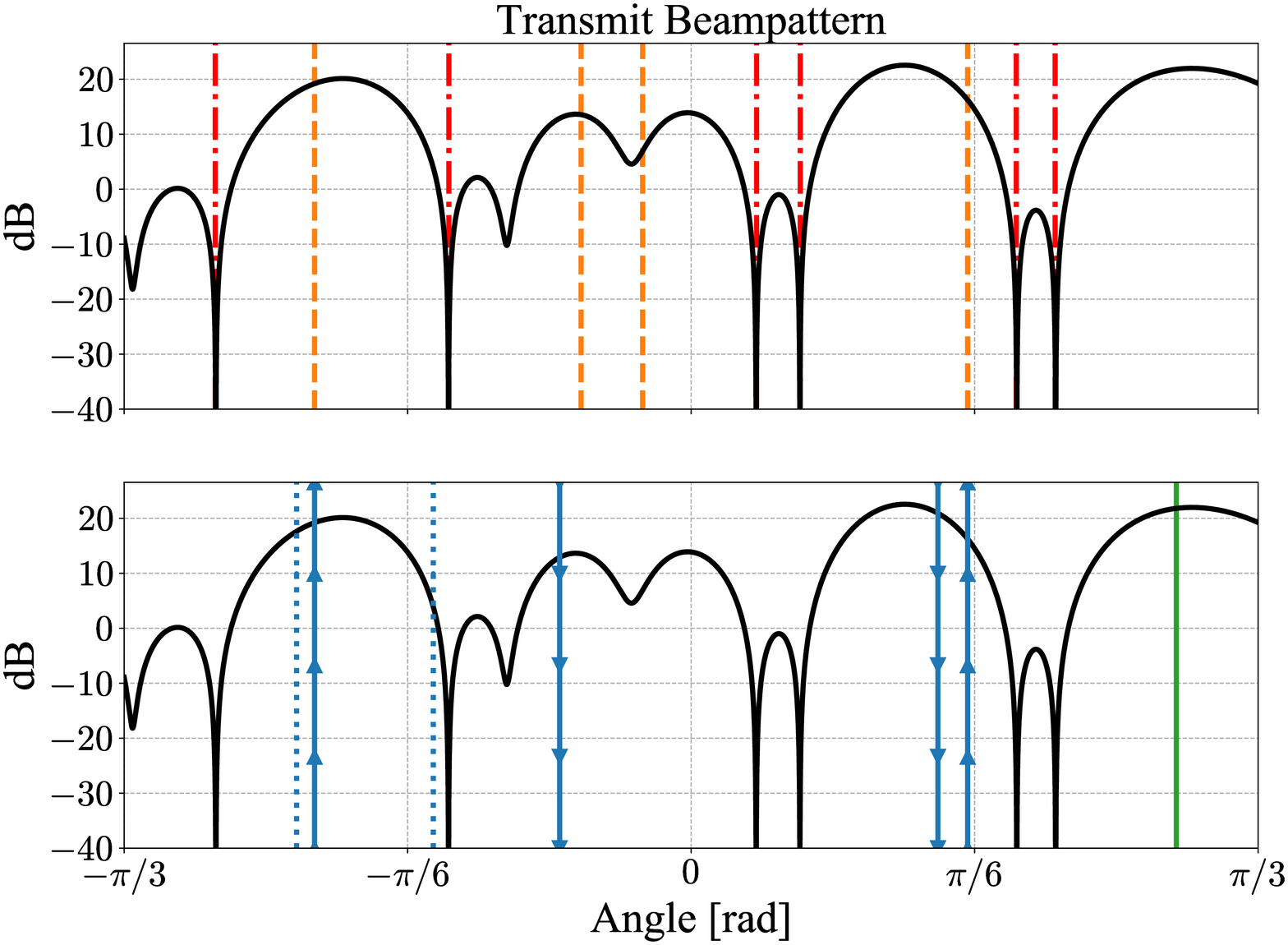}}
		\vspace{-0.2cm}
		\caption{Transmit beampattern (black, solid) {on one subcarrier for one problem instance included in the top-left plot of Fig.~\ref{fig:SINR}, when $\epsilon = 10^{-5}$,  $\delta = 10^{-6}$, and $\text{SCR}=-20$~dB}. Here, the top plot shows the transmit beampattern along with the clutter (orange, dashed) and protected (red, dash dot) directions, while the bottom plot shows the transmit beampattern along with the target (green, solid) and user (blue) directions. There are two direct (blue, dotted) and two indirect (blue, solid) users with two indirect paths each; the indirect users are distinguished by the markers.\label{subfig:radbp}}

		\centerline{\includegraphics[width=1\columnwidth]{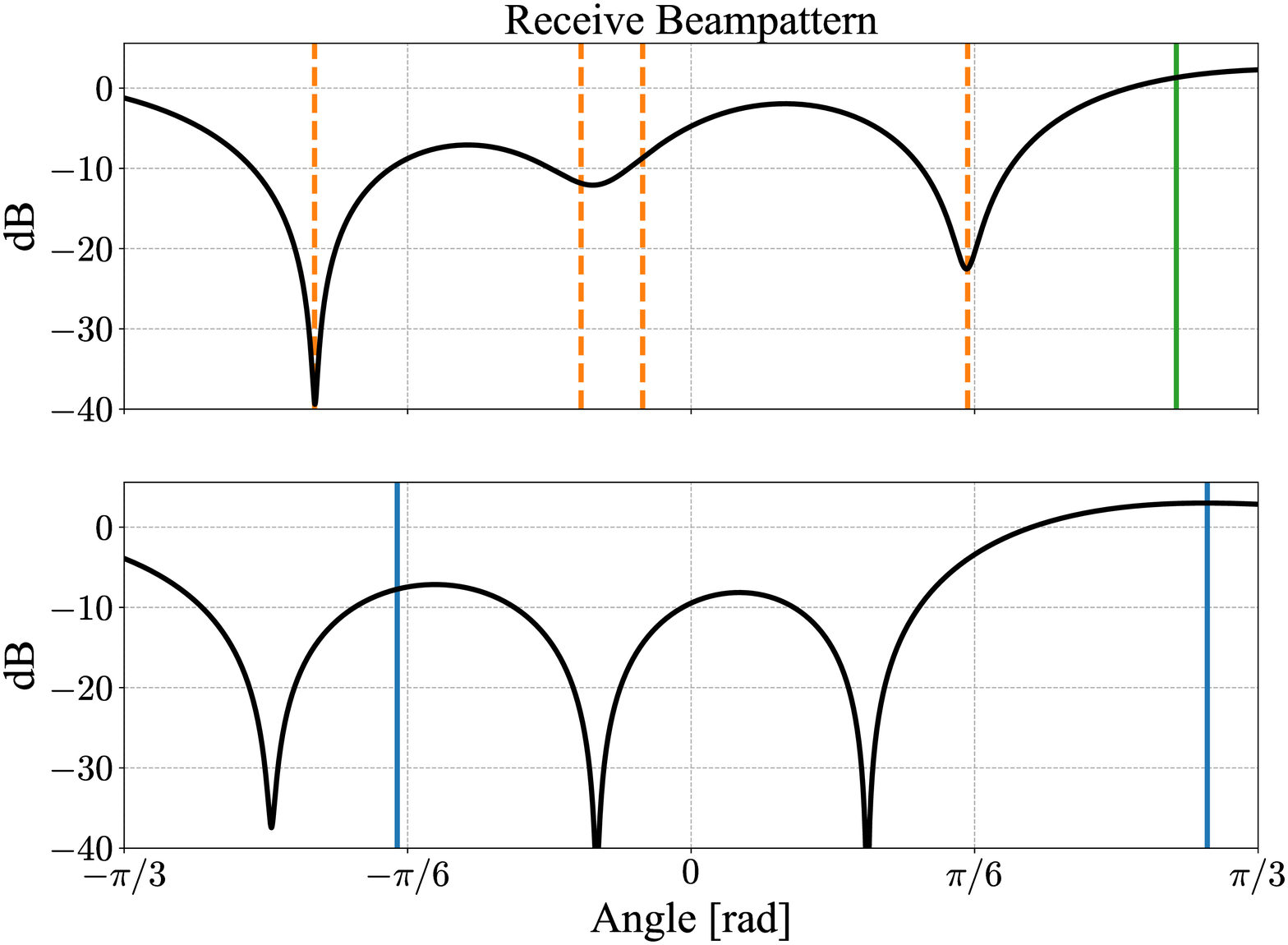}}
		\vspace{-0.2cm}
		\caption{Receive beampatterns (black, solid) {on one subcarrier for one problem instance included  in the top-left plot of Fig.~\ref{fig:SINR}, when $\epsilon = 10^{-5}$,  $\delta = 10^{-6}$, and $\text{SCR}=-20$~dB}. The top plot shows the radar receive beampattern along with the target (green, solid) and clutter (orange, dashed) directions; the bottom plot shows to the receive beampattern of one indirect user along with its receive directions  (blue, solid).\label{subfig:combp}}		
	\end{figure}
	
		\begin{figure}[t]
		\centering
		\centerline{\includegraphics[width=1\columnwidth]{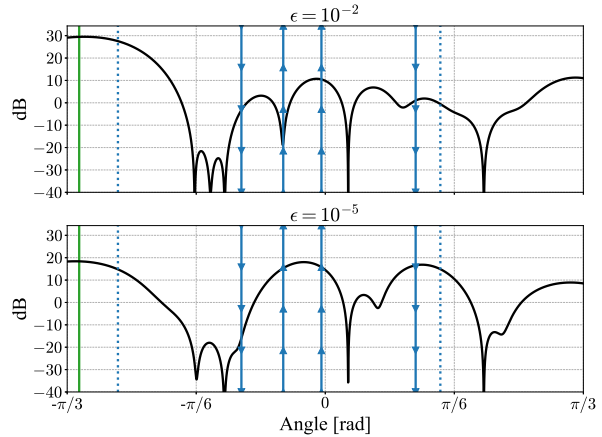}}
		\vspace{-0.2cm}
		\caption{Transmit beampattern (black, solid) {on one subcarrier for one problem instance in the top-left plot of Fig.~\ref{fig:SINR}, when $\epsilon=10^{-2},\,10^{-5}$, $\delta=10^{-6}$ and $\text{SCR}=-20$~dB.} The target (green, solid) and user (blue) directions are superimposed, including two direct (blue, dotted) and two indirect (blue, solid) users with two indirect paths each. The indirect users are distinguished by the markers. }
		\label{fig:bpEmax}
	\end{figure}
		
For one problem instance included in the top-left plot of Fig.~\ref{fig:SINR}, we now visualize the optimized transmit and receive beampatterns in one subcarrier when $\epsilon = 10^{-5}$ and $\delta = 10^{-6}$. The transmit beampattern is defined as in \eqref{eq:txbp}, while the radar and user receive beampatterns are $\| \bm {W}_{k} \herm \bm g_{k}(\xi)\|^2/T$ and $\| \bm {W}_{k,m} \herm \bm g_{k,m}(\xi)\|^2/T$, respectively. Fig.~\ref{subfig:radbp} depicts the transmit beampattern (black, solid); the vertical lines indicate the locations of the clutter (orange, dashed) and protected directions (red, dash dot), in the top plot, and the locations of the radar target (green, solid) and of the direct (blue, dotted) and indirect (blue, solid) users, in the bottom plot. It is verified by inspection that the transmit beampattern peaks at the target location and has nulls at the protected directions. The indirect users are allocated significant power since they require a large SNR to achieve the specified error rate, while it suffices to serve the direct users by sidelobes---the required $\epsilon= 10^{-5}$ corresponds to $\text{SNR}\approx10$~dB for direct users and $\text{SNR}\approx40$~dB for indirect users. For the same scenario considered in Fig.~\ref{subfig:radbp}, Fig.~\ref{subfig:combp} shows (top) the radar receive beampattern with clutter (orange, dashed) and target (green, solid) directions superimposed and (bottom) the receive beampattern of one indirect user with the direction of the received paths superimposed. The radar receive beampattern peaks at the target location and,  together with the transmit beampattern, concurs to mitigate the clutter: indeed, for each clutter direction, we may have a null in the transmit beampattern, a null in the receive beampattern, or sufficiently low values in both beampatterns. The user's receive beampattern instead only emphasizes the two indirect paths.
	
For another problem instance included in the top-left plot of Fig.~\ref{fig:SINR}, Fig.~\ref{fig:bpEmax} shows the effect of decreasing  $\epsilon$ (the constraint on user error rate) on the transmit beampattern obtained in the first subcarrier. Here the line styles match those of Fig.~\ref{subfig:radbp} and $\delta = 10^{-6}$. When $\epsilon = 10^{-2}$, the mainlobe is centered on the target. Instead, when $\epsilon = 10^{-5}$, there are three lobes of similar height directed toward the target and indirect users. Some of the power that was directed toward the radar in the former case has been reallocated to the indirect users in the latter.
    
\begin{figure}[t]
		\centering
		\centerline{\includegraphics[width=0.9\columnwidth]{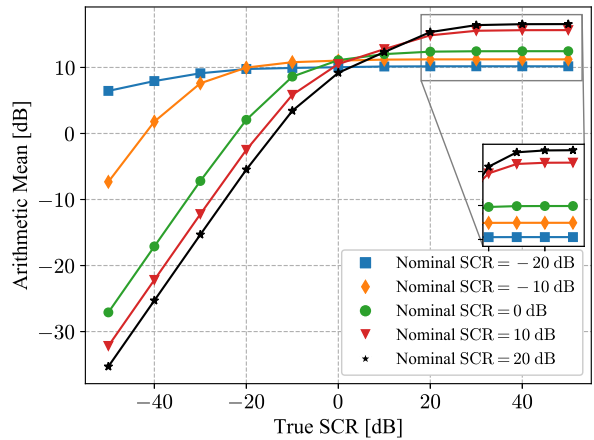}}
		\vspace{-0.2cm}
		\caption{Arithmetic mean of the radar SINRs versus the true SCR, when $\epsilon = 5\times10^{-6}$,   $\delta = 10^{-3}$, and the system is optimized for an SCR of $-20,\,-10,\,0,\,10$, and $20$ dB. \label{fig:evalSCR}}		
	\end{figure}

    We now study the mismatch loss when the true SCR is different from the one employed for  design. Fig.~\ref{fig:evalSCR} reports the arithmetic mean of the radar SINRs versus the true SCR, when {$\epsilon = 10^{-5}$,  $\delta = 10^{-6}$, and} the system is optimized for a nominal SCR of $-20,-10,\,0,\,10,\,20$~dB. Using a nominal SCR of $-20$~dB yields a quite robust design; indeed, the optimization prioritizes nulling the clutter directions, thus making the true strength of the clutter less consequential. When instead the design SCR is $20$~dB, the objective function decays rapidly as the true SCR decreases, while becoming only slightly favored when the true SCR is $20$~dB or greater. 
    \begin{figure}[t]
		\centerline{\includegraphics[width=0.9\columnwidth]{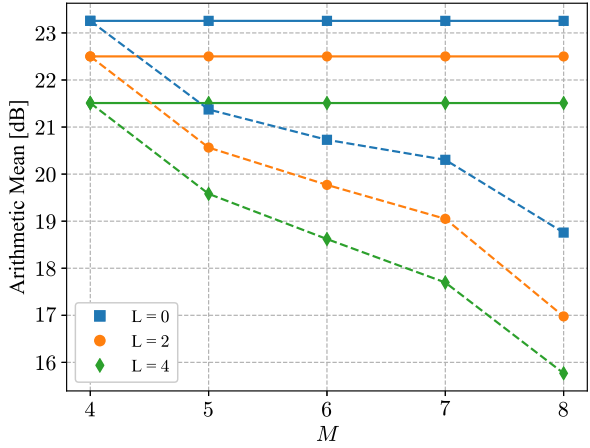}}
		\vspace{-0.2cm}
		\caption{Arithmetic mean of the radar SINRs vs number of users $M$ for $L=0,2,4$ when {$\epsilon = 10^{-5}$,  $\delta = 10^{-4}$, and $\text{SCR}=-20$~dB}. The solid and dashed lines are generated by adding direct and indirect users, respectively.\label{fig:sinrvsM}}
	\end{figure}

    Next, we vary the number of connected users when {$\epsilon = 10^{-5}$,  $\delta = 10^{-4}$, $\text{SCR}=-20$~dB, and} $L\in \{ 0,\,2,\,4\}$. Fig.~\ref{fig:sinrvsM} reports the arithmetic mean of the radar SINRs when there are $4$ direct users and $M$ is increased up to $8$  by adding either indirect users (dashed) or direct users (solid). Adding indirect users causes a severe performance loss; indeed, for the same error rate, the indirect users requires more physical resources than the direct ones to counteract the channel fading.	

\begin{figure}[t]
		\centering
		\centerline{\hspace{20pt}\includegraphics[width=\columnwidth]{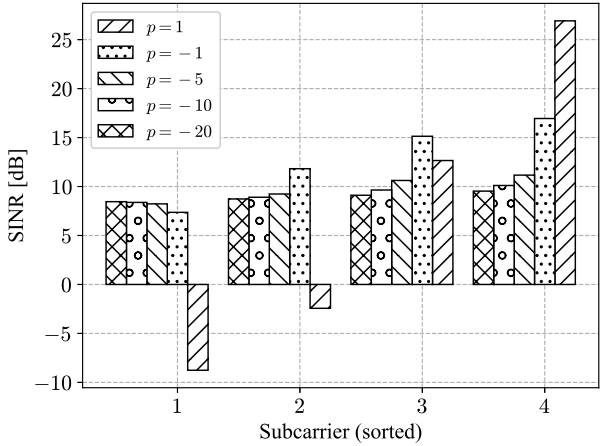}}
		\caption{$p$-power mean of the radar SINRs versus (sorted) subcarrier for $p=-20,-10,-5,-1,1$, when $\epsilon= 4.75\times10^{-5}$, $\delta= 10^{-3}$, and $\text{SCR}=-20$~dB. Here, subcarrier $1$ corresponds to the subcarrier with minimum SINR, and $4$ corresponds to that with maximum SINR.}
		\label{fig:barplot}
	\end{figure}

   Finally, we consider the merit function in~\eqref{p-th-power-mean}, and we assess the effect of changing $p$ on the individual radar SINRs on each subcarrier. For $p\in\{-20,-10,-5,-1,1\}$, we run the Algorithm \ref{proposed_alg} for $10$ instances when $\epsilon= 4.75\times10^{-5}$, $\delta= 10^{-3}$, and $\text{SCR}=-20$~dB; then we compute the average highest SINR obtained by a subcarrier in each instance, the second-highest SINR, and so on. Fig.~\ref{fig:barplot} reports these averages for each $p$. It is seen a smaller $p$ better safeguards the radar operation in the less favorable subcarrier at the price of some performance loss in the most favorable one.
  
	\section{Conclusions}\label{Sec:conclusions}
	In this manuscript, we have considered an OFDM-DFRC system employing a DPSK modulation. We have selected the transmit waveforms and the receive filters  to maximize the radar performance under constraints on the average radiated  power, the error rate of each user, and the beampattern level towards specific directions. The system design results in a non-convex problem, which has been suboptimally solved via an iterative procedure based upon an alternating maximization of the involved variables, a convex restriction of the feasible search set, and the minorization-maximization algorithm. Remarkably, the proposed procedure can be used for a broad family of radar merit functions. The numerical analysis has illustrated the achievable system tradeoffs and the effect of the prior uncertainty on the target strength. 
	
	Future developments may consider the use of reconfigurable intelligent surfaces to reach blind spots or create additional indirect paths, and the use of a differential space-time-frequency code to exploit frequency diversity. Also, the system design when the users simultaneously exploit both direct and indirect signals is an open problem that requires further investigation.
	
	\appendix	
	Here we provide the proofs of Propositions~\ref{proposition_gen_mean},~\ref{Prop-g1},~\ref{Prop-w}, and~\ref{Prop-C3}.
	
	\subsection{Proof of Proposition~\ref{proposition_gen_mean}}\label{proof_gen_mean}	
	Let $y=\sum_{i=1}^K \mu_i \gamma(x_i)$, so that $f(\bm x) = \gamma^{-1}(y)$. Exploiting the formula for the derivative of the inverse function, we have
	\begin{subequations}
		\begin{align}
			\frac{\partial f(\bm x)}{\partial x_i} &= \frac{\mu_i \gamma'(x_i)}{\gamma'\left( \gamma^{-1}(y)\right)}\\
			\frac{\partial^2 f(\bm x)}{\partial x_i^2} &=\frac{\mu_i \gamma''(x_i)}{\gamma'\left( \gamma^{-1}(y)\right)} - \frac{\mu_i^2\left[\gamma'(x_i)\right]^2 \gamma''\left(\gamma^{-1}(y)\right)}{\left[\gamma'\left( \gamma^{-1}(y)\right)\right]^3}\\\
			\frac{\partial^2 f(\bm x)}{\partial x_i \partial x_j} &=  - \frac{\mu_i \mu_j \gamma'(x_i) \gamma'(x_j) \gamma''\left(\gamma^{-1}(y)\right)}{\left[\gamma'\left( \gamma^{-1}(y)\right)\right]^3}
		\end{align}%
	\end{subequations}
	so that the Hessian matrix is
	\begin{align}
		\nabla_f^2(\bm x) &=\frac{1}{\gamma'\left( \gamma^{-1}(y)\right)} \diag\bigl(\{\mu_i \gamma''(x_i)\}\bigr) -\frac{\gamma''\left(\gamma^{-1}(y)\right)}{\left[\gamma'\left( \gamma^{-1}(y)\right)\right]^3}\notag\\
		&\quad \times \begin{pmatrix} \mu_1 \gamma'(x_1) \\ \vdots \\ \mu_K \gamma(x_K) \end{pmatrix}  \begin{pmatrix} \mu_1 \gamma'(x_1) & \cdots & \mu_K \gamma(x_K) \end{pmatrix}
	\end{align}
	that is negative semidefinite for any $\bm x\in \bar{\mathbb R}^K$ if and only if
	\begin{multline}
		\bm z\transp \nabla_f^2(\bm x) \bm z = \frac{1}{\left[\gamma'\left( \gamma^{-1}(y)\right)\right]^3} \Bigg( \left[\gamma'\left( \gamma^{-1}(y)\right)\right]^2 \\
		\times\sum_{i=1}^K \mu_i \gamma''(x_i)z_i^2 - \gamma''\left( \gamma^{-1}(y)\right) \left( \sum_{i=1}^K \mu_i \gamma'(x_i)z_i\right)^2  \Bigg) \leq0 \label{H_neg_semidef}
	\end{multline}
	for all $\bm z\in\mathbb R^K$ and any $\bm x\in\bar{\mathbb R}^K$.
	
	In order to prove~\eqref{H_neg_semidef}, we follow and generalize the approach in~\cite[Ch.~III, Sec.~16]{Hardy_1934}, where the convexity of $f$ (instead of the concavity) is proven under the condition that $\gamma\in C^4$ is strictly positive, strictly increasing, and strictly convex. In particular, since either $\gamma'(x)>0$ and $\gamma''(x)<0$ for all $x\in\bar{\mathbb R}$, or $\gamma'(x)<0$ and $\gamma''(x)>0$ for all $x\in\bar{\mathbb R}$, inequality~\eqref{H_neg_semidef} holds if and only if, for any $\bm x\in\bar{\mathbb R}^K$,
	\begin{equation}
		\frac{\left[\gamma'\left( \gamma^{-1}(y)\right)\right]^2}{|\gamma''\left( \gamma^{-1}(y)\right)|} \geq \frac{\left( \sum_{i=1}^K \mu_i \gamma'(x_i)z_i\right)^2}{\sum_{i=1}^K \mu_i |\gamma''(x_i)| z_i^2}, \quad \forall \bm z\in\mathbb R^K. \label{concave_cond_1}
	\end{equation}
	Now, by the Cauchy-Schwarz inequality, we have that
	\begin{align}
		\left( \sum_{i=1}^K \mu_i \gamma'(x_i)z_i\right)^2 &= \left(\sum_{i=1}^K z_i \sqrt{ \mu_i |\gamma''(x_i)|} \sqrt{\frac{\mu_i \left[\gamma'(x_i)\right]^2}{|\gamma''(x_i)|}}\right)^2\notag\\
		&\leq \sum_{i=1}^K \mu_i |\gamma''(x_i)| z_i^2 \sum_{i=1}^K \mu_i \frac{ \left[\gamma'(x_i)\right]^2}{|\gamma''(x_i)|}
	\end{align}
	with equality if and only if $z_i$ is proportional to $\frac{\gamma'(x_i)}{|\gamma''(x_i)|}$. Therefore, condition~\eqref{concave_cond_1} holds if and only if
	\begin{equation}
		\frac{\left[\gamma'\left( \gamma^{-1}(y)\right)\right]^2}{|\gamma''( \gamma^{-1}(y))|} \geq \sum_{i=1}^K \mu_i \frac{ \left[\gamma'(x_i)\right]^2}{|\gamma''(x_i)|}, \quad \forall \bm x \label{concave_cond_2}
	\end{equation}
	which, upon defining $
		g(y)=\left[\gamma'\left( \gamma^{-1}(y)\right)\right]^2/ \big[|\gamma''( \gamma^{-1}(y))|\big]$
	and recalling that $y=\sum_{i=1}^K \mu_i \gamma(x_i)$, becomes
	\begin{equation}
		g\left( \sum_{i=1}^K \mu_i \gamma(x_i) \right) \geq \sum_{i=1}^K \mu_i g(x_i), \quad \forall \bm x \in\bar{\mathbb R}^K. \label{concave_cond_3}
	\end{equation}
	This is a concavity condition on $g$ that, since $\gamma \in C^4$, is satisfied if and only if $g''(y)\leq 0$, for all $y$. Finally, since
	\begin{subequations}
		\begin{align}
			g'(y)&= \left. -\frac{d}{dx} \frac{\gamma'(x)}{\gamma''(x)}\right|_{x=\gamma^{-1}(y)} -1 \\
			g''(y)&= -\left. \frac{1}{\gamma'(x)} \frac{d^2}{dx^2} \frac{\gamma'(x)}{\gamma''(x)}\right|_{x=\gamma^{-1}(y)}
		\end{align}%
	\end{subequations}
	if $\gamma'(x)>0$ and $\gamma''(x)<0$ for all $x\in\bar{\mathbb R}$, and
	\begin{subequations}
		\begin{align}
			g'(y)&= \left. \frac{d}{dx} \frac{\gamma'(x)}{\gamma''(x)}\right|_{x=\gamma^{-1}(y)} +1 \\
			g''(y)&= \left. \frac{1}{\gamma'(x)} \frac{d^2}{dx^2} \frac{\gamma'(x)}{\gamma''(x)}\right|_{x=\gamma^{-1}(y)}
		\end{align}%
	\end{subequations}
	if $\gamma'(x)<0$ and $\gamma''(x)>0$ for all $x\in\bar{\mathbb R}$, we have that $g$ is concave if and only if $\gamma'/\gamma''$ is convex.
	
	\subsection{Proof of Proposition~\ref{Prop-g1}}\label{proof_Prop-g1}
	The function $g$ is twice-differentiable with 
	\begin{subequations}
		\begin{align}
		    \nabla_{g}(\bm u,x)&= \begin{pmatrix} \frac{2}{x} \bm \varPsi \bm u\\
			-\frac{1}{x^2} \bm u\herm\bm \varPsi \bm u
			\end{pmatrix} \label{derivatives_g1_g1}\\
			\nabla_{g}^2(\bm u,x) & =\frac{2}{x}\begin{pmatrix}
			\bm{\varPsi} & -\frac{1}{x}\bm{\varPsi} \bm{u}\\
			-\frac{1}{x}\bm{u}\herm\bm{\varPsi}  & \hspace{5pt}\frac{1}{x^{2}}\bm{u}\herm\bm{\varPsi}\bm{u}	
		\end{pmatrix}.
		\end{align}
	\end{subequations}
	Since $\bm \varPsi\succeq0$ and 
	\begin{subequations}
		\begin{align}
			& (\bm I_{N} - \bm \varPsi \bm \varPsi^\dagger)\left(-\frac{1}{x} \bm{\varPsi} \bm u\right)= \bm 0_{N}\\
			& \frac{1}{x^2}\bm{u}\herm\bm{\varPsi} \bm u - \left(\frac{1}{x}\bm{u}\herm\bm{\varPsi}\right) \bm \varPsi^\dagger \left( \frac{1}{x} \bm{\varPsi} \bm u\right) = 0
		\end{align}
	\end{subequations}
	we conclude that $\nabla_{g}^2(\bm u,x)\succeq 0$ for any $\bm u \in \mathbb{C}^{N}$ and $x >0$~\cite[Sec.~A.5.5]{boyd2004convex}, so that $g$ is convex. 
	
	\subsection{Proof of Proposition~\ref{Prop-w}}\label{proof_Prop-w}	
	The filter $\bm{w}_{k,m}$ only comes into play in C3 and C4.  
	Assume first that $\kappa_{k,m}=\infty$. For any feasible $\bm{u}_{k}$, the optimal $\bm{w}_{k,m}$ must maximize $\text{SNR}_{k,m}$ under the constraint set $\mathcal{W}_{k,m}^{\rm d}=
		\big\{\bm{w}_{k,m}\in \mathbb{C}^{TN_{m}}:\,\kappa_{k,m}=\infty\big\}$.
	Notice now that
	\begin{align}
		&\max_{\bm{w}_{k,m}\in\mathcal{W}_{k,m}^{\rm d}}\text{SNR}_{k,m}=\max_{\bm{w}_{k,m}\in\mathcal{W}_{k,m}^{\rm d}}\frac{\bm{w}_{k,m}\herm \bm{\Xi}_{k,m}	\bm{w}_{k,m} }{\|\bm{w}_{k,m}\|^2}\notag \\
		&=\max_{\bm{w}_{k,m}\in\mathcal{W}_{k,m}^{\rm d}}\frac{\bm{w}_{k,m}\herm \bm{\Xi}_{k,m}^{\rm d}	\bm{w}_{k,m} }{\|\bm{w}_{k,m}\|^2}\leq \lambda_{\max}(\bm{\Xi}_{k,m}^{\rm d}).
	\end{align}
	In the above derivations, the first equality follows from~\eqref{eq: SINR-kappa-1},  \eqref{SNR-def}, and~\eqref{Xi-def}; the second equality is a consequence of the fact that $\big(\bm{I}_{T}\otimes\bm{\Pi}_{k,m}^{\rm d}\big)\bm{w}_{k,m}=\bm{w}_{k,m}$ if $\bm{w}_{k,m}\in\mathcal{W}_{k,m}^{\rm d}$; finally, the last inequality is tight when  $\bm{w}_{k,m}$ is proportional to the eigenvector  corresponding to the largest eigenvalue of $\bm{\Xi}_{k,m}^{\rm d}$.	The result for $\kappa_{k,m}=0$ follows by similar arguments.
	
	\subsection{Proof of Proposition~\ref{Prop-C3}}\label{proof_Prop-C3}	
	Upon solving for the optimal $\bm{w}_{k,m}$ as reported in Proposition~\ref{Prop-w}, the constraint C3 can be reformulated as 
	\begin{equation}
		\begin{cases}
			\lambda_{\max}(\bm{\Xi}_{k,m}^{\rm d}) \geq \rho_{k,m}^{\rm d},& \text{if } \kappa_{k,m}=\infty\\
			\lambda_{\max}(\bm{\Xi}_{k,m}^{\rm i}) \geq \rho_{k,m}^{\rm i},& \text{if } \kappa_{k,m}=0
		\end{cases},\quad  \forall k,m.
	\end{equation}
	If $\kappa_{k,m}=\infty$, then we have
	\begin{align}
		\lambda_{\max}(\bm{\Xi}_{k,m}^{\rm d})&=\frac{|\beta_{k,m,0}|^2}{\sigma^2_{z,k,m}}\big\|\left(\bm{I}_{T}\otimes\bm{\Pi}_{k,m}^{\rm d}\right)\notag\\ 
		& \quad \times 	\left(\bm{I}_{T}\otimes\bm{g}_{k,m}(\bar \phi_{m,0})\bm{s}_{k}\transp( \phi_{m,0})\right)\bm{u}_{k}\big\|^2\notag \\
		&= \frac{|\beta_{k,m,0}|^2}{\sigma^2_{z,k,m}} \bm{g}_{k,m}\herm(\bar \phi_{m,0})\bm{\Pi}_{k,m}^{\rm d}\bm{g}_{k,m}(\bar \phi_{m,0}) \notag\\ 
		& \quad \times \bm{u}_{k}\herm \left(\bm{I}_{T}\otimes\bm{s}_{k}^{*}(\phi_{m,0})\bm{s}_{k}\transp( \phi_{m,0})\right)\bm{u}_{k}\notag \\ 
		&= \frac{|\beta_{k,m,0}|^2}{\sigma^2_{z,k,m}} \bm{g}_{k,m}\herm(\bar \phi_{m,0})\bm{\Pi}_{k,m}^{\rm d}\bm{g}_{k,m}(\bar \phi_{m,0}) \notag\\ 
		& \quad \times  \trace\{\bm{U}_{k}\herm\bm{s}_{k}^{*}(\phi_{m,0})\bm{s}_{k}\transp( \phi_{m,0})\bm{U}_{k}\}.
	\end{align} 
	The above derivations show that we must necessarily have $\bm{U}_{k}\herm\bm{s}_{k}^{*}(\phi_{m,0})\neq \bm{0}_{N_{t}}$, as otherwise $\lambda_{\max}(\bm{\Xi}_{k,m})=0$. Also, if $\text{rank}\{\bm{U}_{k}\}=N_{t}$, we can write~\cite{Horn-2012}
	\begin{align}
		\lambda_{\max}(\bm{\Xi}_{k,m}^{\rm d})&\geq 
		\frac{N_{t}|\beta_{k,m,0}|^2}{\sigma^2_{z,k,m}} \lambda_{\min}(\bm{U}_{k}\bm{U}_{k}\herm) \notag \\ &\quad \times \bm{g}_{k,m}\herm(\bar \phi_{m,0})\bm{\Pi}_{k,m}^{\rm d}\bm{g}_{k,m}(\bar \phi_{m,0})   
	\end{align} 
	and C3 can be satisfied if~\eqref{C3_feasibility} holds.
	
	If $\kappa_{k,m}=0$, from the Weyl's Theorem~\cite{Horn-2012} we have
	\begin{subequations}
	\begin{align}
		\lambda_{\max}(\bm{\Xi}_{k,m}^{\rm i}) &\leq 		\sum_{q=1}^{Q_{m}}\dfrac{\sigma_{\beta,k,m,q}^2}{\sigma^2_{z,k,m}} \bm{g}_{k,m}\herm(\bar \phi_{m,q})\bm{\Pi}_{k,m}^{\rm i}\bm{g}_{k,m}(\bar \phi_{m,q}) \notag \\ &\quad \times \trace\{\bm{U}_{k}\herm\bm{s}_{k}^{*}(\phi_{m,q})\bm{s}_{k}\transp( \phi_{m,q})\bm{U}_{k}\}\\	
		\lambda_{\max}(\bm{\Xi}_{k,m}^{\rm i}) &\geq\max_{q}\dfrac{\sigma_{\beta,k,m,q}^2}{\sigma^2_{z,k,m}} \bm{g}_{k,m}\herm(\bar \phi_{m,q})\bm{\Pi}_{k,m}^{\rm i}\bm{g}_{k,m}(\bar \phi_{m,q}) \notag \\ &\quad \times \trace\{\bm{U}_{k}\herm\bm{s}_{k}^{*}(\phi_{m,q})\bm{s}_{k}\transp( \phi_{m,q})\bm{U}_{k}\}.
	\end{align}
	\end{subequations}
	The above inequalities show that we must necessarily have $\bm{U}_{k}\herm \bm{s}_{k}^{*}( \phi_{m,q})\neq \bm{0}_{N_{t}}$ for at least one indirect path $q\in\{1,\ldots,Q_{m}\}$.
	Also, if $\text{rank}\{\bm{U}_{k}\}=N_{t}$, we can write~\cite{Horn-2012}
	\begin{align}
		\lambda_{\max}(\bm{\Xi}_{k,m}^{\rm i})&\geq \max_{q}\dfrac{N_{t}\sigma_{\beta,k,m,q}^2}{\sigma^2_{z,k,m}}\lambda_{\min}(\bm{U}_{k}\bm{U}_{k}\herm)\notag \\ &\quad \times \bm{g}_{k,m}\herm(\bar \phi_{m,q})\bm{\Pi}_{k,m}^{\rm i}\bm{g}_{k,m}(\bar \phi_{m,q}) 
	\end{align} 
	and C3 can be satisfied if~\eqref{C3_feasibility} holds.


\begin{thebibliography}{10}
	\providecommand{\url}[1]{#1}
	\csname url@samestyle\endcsname
	\providecommand{\newblock}{\relax}
	\providecommand{\bibinfo}[2]{#2}
	\providecommand{\BIBentrySTDinterwordspacing}{\spaceskip=0pt\relax}
	\providecommand{\BIBentryALTinterwordstretchfactor}{4}
	\providecommand{\BIBentryALTinterwordspacing}{\spaceskip=\fontdimen2\font plus
		\BIBentryALTinterwordstretchfactor\fontdimen3\font minus
		\fontdimen4\font\relax}
	\providecommand{\BIBforeignlanguage}[2]{{%
			\expandafter\ifx\csname l@#1\endcsname\relax
			\typeout{** WARNING: IEEEtran.bst: No hyphenation pattern has been}%
			\typeout{** loaded for the language `#1'. Using the pattern for}%
			\typeout{** the default language instead.}%
			\else
			\language=\csname l@#1\endcsname
			\fi
			#2}}
	\providecommand{\BIBdecl}{\relax}
	\BIBdecl
	\renewcommand{\BIBentryALTinterwordstretchfactor}{4}
	
	\bibitem{Griffiths-2015}
	H.~Griffiths \emph{et~al.}, ``Radar spectrum engineering and management:
	Technical and regulatory issues,'' \emph{Proceedings of the IEEE}, vol. 103,
	no.~1, pp. 85--102, Jan. 2015.
	
	\bibitem{mazar2016radio}
	H.~Mazar, \emph{Radio spectrum Management: Policies, regulations and
		techniques}.\hskip 1em plus 0.5em minus 0.4em\relax {USA}: John Wiley \&
	Sons, 2016.
	
	\bibitem{Tarokh-2008}
	N.~Devroye, M.~Vu, and V.~Tarokh, ``Cognitive radio networks,'' \emph{IEEE
		Signal Processing Magazine}, vol.~25, no.~6, pp. 12--23, Nov. 2008.
	
	\bibitem{Proakis-book}
	J.~Proakis and M.~Salehi, \emph{Digital Communications}, 5th~ed.\hskip 1em plus
	0.5em minus 0.4em\relax New York, NY, USA: McGraw-Hill Higher Education,
	2014.
	
	\bibitem{Hanzo-2018}
	L.~Dai \emph{et~al.}, ``A survey of non-orthogonal multiple access for {5G},''
	\emph{IEEE Communications Surveys Tutorials}, vol.~20, no.~3, pp. 2294--2323,
	thirdquarter 2018.
	
	\bibitem{Venturino2009}
	L.~Venturino, N.~Prasad, and X.~Wang, ``Coordinated scheduling and power
	allocation in downlink multicell {OFDMA} networks,'' \emph{IEEE Transactions
		on Vehicular Technology}, vol.~58, no.~6, pp. 2835--2848, Jul. 2009.
	
	\bibitem{Venturino2011}
	H.~Zhang \emph{et~al.}, ``Weighted sum-rate maximization in multi-cell networks
	via coordinated scheduling and discrete power control,'' \emph{IEEE Journal
		on Selected Areas in Communications}, vol.~29, no.~6, pp. 1214--1224, Jun.
	2011.
	
	\bibitem{Sanguinetti-2020}
	E.~Björnson and L.~Sanguinetti, ``Scalable cell-free massive {MIMO} systems,''
	\emph{IEEE Transactions on Communications}, vol.~68, no.~7, pp. 4247--4261,
	Jul. 2020.
	
	\bibitem{Venturino-2007}
	L.~Venturino, N.~Prasad, X.~Wang, and M.~Madihian, ``Design of linear
	dispersion codes for practical {MIMO-OFDM} systems,'' \emph{IEEE Journal of
		Selected Topics in Signal Processing}, vol.~1, no.~1, pp. 178--188, Jun.
	2007.
	
	\bibitem{Fatema-2018}
	N.~Fatema, G.~Hua, Y.~Xiang, D.~Peng, and I.~Natgunanathan, ``Massive {MIMO}
	linear precoding: A survey,'' \emph{IEEE Systems Journal}, vol.~12, no.~4,
	pp. 3920--3931, Dec. 2018.
	
	\bibitem{Albreem-2019}
	M.~A. Albreem, M.~Juntti, and S.~Shahabuddin, ``Massive {MIMO} detection
	techniques: A survey,'' \emph{IEEE Communications Surveys Tutorials},
	vol.~21, no.~4, pp. 3109--3132, Fourthquarter 2019.
	
	\bibitem{Farina2020}
	S.~H. Javadi and A.~Farina, ``Radar networks: A review of features and
	challenges,'' \emph{Information Fusion}, vol.~61, pp. 48--55, 2020.
	
	\bibitem{Griffiths-2006}
	S.~Miranda, C.~Baker, K.~Woodbridge, and H.~Griffiths, ``Knowledge-based
	resource management for multifunction radar: a look at scheduling and task
	prioritization,'' \emph{IEEE Signal Processing Magazine}, vol.~23, no.~1, pp.
	66--76, Jan. 2006.
	
	\bibitem{Griffiths-2019}
	S.~Z. Gurbuz \emph{et~al.}, ``An overview of cognitive radar: Past, present,
	and future,'' \emph{IEEE Aerospace and Electronic Systems Magazine}, vol.~34,
	no.~12, pp. 6--18, Dec. 2019.
	
	\bibitem{Li_Stoica_2009}
	J.~Li and P.~Stoica, \emph{{MIMO} radar signal processing}.\hskip 1em plus
	0.5em minus 0.4em\relax Hoboken, {USA}: John Wiley \& Sons, 2009.
	
	\bibitem{Mokole-2016}
	S.~D. Blunt and E.~L. Mokole, ``Overview of radar waveform diversity,''
	\emph{IEEE Aerospace and Electronic Systems Magazine}, vol.~31, no.~11, pp.
	2--42, Nov. 2016.
	
	\bibitem{Venturino-MIMO-2011}
	E.~Grossi, M.~Lops, and L.~Venturino, ``Robust waveform design for{ MIMO}
	radars,'' \emph{IEEE Transactions on Signal Processing}, vol.~59, no.~7, pp.
	3262--3271, Jul. 2011.
	
	\bibitem{Venturino-MIMO-2012}
	E.~Grossi, M.~Lops, and L.~Venturino, ``Min–max waveform design for {MIMO}
	radars under unknown correlation of the target scattering,'' \emph{Signal
		Processing}, vol.~92, no.~6, pp. 1550--1558, 2012.
	
	\bibitem{Stoica-2010}
	W.~Roberts, P.~Stoica, J.~Li, T.~Yardibi, and F.~A. Sadjadi, ``Iterative
	adaptive approaches to {MIMO} radar imaging,'' \emph{IEEE Journal of Selected
		Topics in Signal Processing}, vol.~4, no.~1, pp. 5--20, Feb 2010.
	
	\bibitem{Li-2019}
	J.~Liu and J.~Li, ``Robust detection in {MIMO} radar with steering vector
	mismatches,'' \emph{IEEE Transactions on Signal Processing}, vol.~67, no.~20,
	pp. 5270--5280, Oct. 2019.
	
	\bibitem{Rangaswamy-2019}
	K.~Alhujaili, V.~Monga, and M.~Rangaswamy, ``Transmit {MIMO} radar beampattern
	design via optimization on the complex circle manifold,'' \emph{IEEE
		Transactions on Signal Processing}, vol.~67, no.~13, pp. 3561--3575, Jul.
	2019.
	
	\bibitem{Cui-2019}
	X.~Yu, G.~Cui, J.~Yang, L.~Kong, and J.~Li, ``Wideband {MIMO} radar waveform
	design,'' \emph{IEEE Transactions on Signal Processing}, vol.~67, no.~13, pp.
	3487--3501, Jul. 2019.
	
	\bibitem{Lops-2019}
	L.~Zheng, M.~Lops, Y.~C. Eldar, and X.~Wang, ``Radar and communication
	coexistence: An overview: A review of recent methods,'' \emph{IEEE Signal
		Processing Magazine}, vol.~36, no.~5, pp. 85--99, Sep. 2019.
	
	\bibitem{Liu-2020}
	F.~{Liu}, C.~{Masouros}, A.~{Petropulu}, H.~{Griffiths}, and L.~{Hanzo},
	``Joint radar and communication design: Applications, state-of-the-art, and
	the road ahead,'' \emph{IEEE Transactions on Communications}, vol.~68, no.~6,
	pp. 3834 -- 3862, Jun. 2020.
	
	\bibitem{Kim-2021}
	N.~C. Luong, X.~Lu, D.~T. Hoang, D.~Niyato, and D.~I. Kim, ``Radio resource
	management in joint radar and communication: A comprehensive survey,''
	\emph{IEEE Communications Surveys Tutorials}, vol.~23, no.~2, pp. 780--814,
	2021.
	
	\bibitem{Hongbin-2019}
	F.~Wang, H.~Li, and M.~A. Govoni, ``Power allocation and co-design of
	multicarrier communication and radar systems for spectral coexistence,''
	\emph{IEEE Transactions on Signal Processing}, vol.~67, no.~14, pp.
	3818--3831, Jul. 2019.
	
	\bibitem{Liao-2019}
	Z.~Cheng, B.~Liao, S.~Shi, Z.~He, and J.~Li, ``Co-design for overlaid {MIMO}
	radar and downlink {MISO} communication systems via {Cramér–Rao} bound
	minimization,'' \emph{IEEE Transactions on Signal Processing}, vol.~67,
	no.~24, pp. 6227--6240, Dec. 2019.
	
	\bibitem{Grossi-2020}
	E.~Grossi, M.~Lops, and L.~Venturino, ``Joint design of surveillance radar and
	{MIMO} communication in cluttered environments,'' \emph{IEEE Transactions on
		Signal Processing}, vol.~68, pp. 1544--1557, 2020.
	
	\bibitem{Grossi-2021}
	E.~Grossi, M.~Lops, and L.~Venturino, ``Energy efficiency optimization in
	radar-communication spectrum sharing,'' \emph{IEEE Transactions on Signal
		Processing}, vol.~69, pp. 3541--3554, 2021.
	
	\bibitem{Qian-TSP-2021}
	J.~Qian, L.~Venturino, M.~Lops, and X.~Wang, ``Radar and communication spectral
	coexistence in range-dependent interference,'' \emph{IEEE Transactions on
		Signal Processing}, vol.~69, pp. 5891--5906, 2021.
	
	\bibitem{Grossi-TWC-2020}
	E.~{Grossi}, M.~{Lops}, and L.~{Venturino}, ``Adaptive detection and
	localization exploiting the {IEEE 802.11ad} standard,'' \emph{IEEE
		Transactions on Wireless Communications}, vol.~19, no.~7, pp. 4394--4407,
	Jul. 2020.
	
	\bibitem{Grossi-TWC-2021}
	E.~Grossi, M.~Lops, A.~M. Tulino, and L.~Venturino, ``Opportunistic sensing
	using mmwave communication signals: A subspace approach,'' \emph{IEEE
		Transactions on Wireless Communications}, vol.~20, no.~7, pp. 4420--4434, jul
	2021.
	
	\bibitem{Hassanien-2016}
	A.~Hassanien, M.~G. Amin, Y.~D. Zhang, and F.~Ahmad, ``Signaling strategies for
	dual-function radar communications: an overview,'' \emph{IEEE Aerospace and
		Electronic Systems Magazine}, vol.~31, no.~10, pp. 36--45, Oct. 2016.
	
	\bibitem{Zhang-2018}
	A.~Ahmed, Y.~D. Zhang, and Y.~Gu, ``Dual-function radar-communications using
	{QAM}-based sidelobe modulation,'' \emph{Digital Signal Processing}, vol.~82,
	pp. 166--174, 2018.
	
	\bibitem{Shlezinger-2020}
	T.~Huang, N.~Shlezinger, X.~Xu, Y.~Liu, and Y.~C. Eldar, ``{MAJoRCom: A}
	dual-function radar communication system using index modulation,'' \emph{IEEE
		Transactions on Signal Processing}, vol.~68, pp. 3423--3438, 2020.
	
	\bibitem{Sturm-2011}
	C.~{Sturm} and W.~{Wiesbeck}, ``Waveform design and signal processing aspects
	for fusion of wireless communications and radar sensing,'' \emph{Proceedings
		of the IEEE}, vol.~99, no.~7, pp. 1236--1259, Jul. 2011.
	
	\bibitem{Dingyou-2020}
	D.~Ma, N.~Shlezinger, T.~Huang, Y.~Liu, and Y.~C. Eldar, ``Joint
	radar-communication strategies for autonomous vehicles: Combining two key
	automotive technologies,'' \emph{IEEE Signal Processing Magazine}, vol.~37,
	no.~4, pp. 85--97, Jul. 2020.
	
	\bibitem{Koivunen-2016}
	M.~{Bică} and V.~{Koivunen}, ``Generalized multicarrier radar: Models and
	performance,'' \emph{IEEE Transactions on Signal Processing}, vol.~64,
	no.~17, pp. 4389--4402, Sep. 2016.
	
	\bibitem{Hassanien-2019}
	A.~Ahmed, Y.~D. Zhang, A.~Hassanien, and B.~Himed, ``{OFDM}-based joint
	radar-communication system: Optimal sub-carrier allocation and power
	distribution by exploiting mutual information,'' in \emph{2019 53rd Asilomar
		Conference on Signals, Systems, and Computers}, Nov. 2019, pp. 559--563.
	
	\bibitem{Shi-2021}
	C.~Shi, Y.~Wang, F.~Wang, S.~Salous, and J.~Zhou, ``Joint optimization scheme
	for subcarrier selection and power allocation in multicarrier dual-function
	radar-communication system,'' \emph{IEEE Systems Journal}, vol.~15, no.~1,
	pp. 947--958, Mar. 2021.
	
	\bibitem{Koivunen-2019}
	M.~Bică and V.~Koivunen, ``Multicarrier radar-communications waveform design
	for {RF} convergence and coexistence,'' in \emph{ICASSP 2019 - 2019 IEEE
		International Conference on Acoustics, Speech and Signal Processing
		(ICASSP)}, May 2019, pp. 7780--7784.
	
	\bibitem{Temiz-2020}
	M.~Temiz, E.~Alsusa, and M.~W. Baidas, ``A dual-functional massive {MIMO OFDM}
	communication and radar transmitter architecture,'' \emph{IEEE Transactions
		on Vehicular Technology}, vol.~69, no.~12, pp. 14\,974--14\,988, Dec. 2020.
	
	\bibitem{Temiz-2021}
	M.~Temiz, E.~Alsusa, and M.~W. Baidas, ``Optimized precoders for massive {MIMO
		OFDM} dual radar-communication systems,'' \emph{IEEE Transactions on
		Communications}, vol.~69, no.~7, pp. 4781--4794, Jul. 2021.
	
	\bibitem{Temiz-2021-uplink}
	M.~Temiz, E.~Alsusa, and M.~W. Baidas, ``A dual-function massive {MIMO} uplink
	{OFDM} communication and radar architecture,'' \emph{IEEE Transactions on
		Cognitive Communications and Networking}, 2021.
	
	\bibitem{Hardy_1934}
	G.~H. Hardy, J.~E. Littlewood, and G.~P\'olya, \emph{Inequalities}.\hskip 1em
	plus 0.5em minus 0.4em\relax Cambridge, {UK}: Cambridge University Press,
	1934.
	
	\bibitem{Handbook-Means-2003}
	P.~Bullen, \emph{Handbook of Means and Their Inequalities}, ser. Mathematics
	and Its Applications.\hskip 1em plus 0.5em minus 0.4em\relax Netherlands:
	Springer, 2003.
	
	\bibitem{Lee-2017}
	H.~Lee and S.~Kim, ``{Muirhead’s and Holland’s} inequalities of mixed power
	means for positive real numbers,'' \emph{Journal of applied mathematics \&
		informatics}, vol.~35, no. 1-2, p. 33–44, Jan. 2017.
	
	\bibitem{Blum-2007}
	Y.~{Yang} and R.~S. {Blum}, ``{MIMO} radar waveform design based on mutual
	information and minimum mean-square error estimation,'' \emph{IEEE
		Transactions on Aerospace and Electronic Systems}, vol.~43, no.~1, pp.
	330--343, Jan. 2007.
	
	\bibitem{Venturino-MIMO-2008}
	A.~De~Maio, M.~Lops, and L.~Venturino, ``Diversity-integration tradeoffs in
	{MIMO} detection,'' \emph{IEEE Transactions on Signal Processing}, vol.~56,
	no.~10, pp. 5051--5061, Oct. 2008.
	
	\bibitem{Nehorai-2010}
	S.~{Sen} and A.~{Nehorai}, ``{OFDM MIMO} radar with mutual-information waveform
	design for low-grazing angle tracking,'' \emph{IEEE Transactions on Signal
		Processing}, vol.~58, no.~6, pp. 3152--3162, Jun. 2010.
	
	\bibitem{Venturino-MIMO-2010}
	A.~Aubry, M.~Lops, A.~M. Tulino, and L.~Venturino, ``On {MIMO} detection under
	non-{Gaussian} target scattering,'' \emph{IEEE Transactions on Information
		Theory}, vol.~56, no.~11, pp. 5822--5838, Nov. 2010.
	
	\bibitem{Jajamovich_2010}
	G.~H. Jajamovich, M.~Lops, and X.~Wang, ``Space-time coding for {MIMO} radar
	detection and ranging,'' \emph{IEEE Transactions on Signal Processing},
	vol.~58, no.~12, pp. 6195--6206, Dec. 2010.
	
	\bibitem{Richards_2005}
	M.~A. Richards, \emph{Fundamentals of radar signal processing}.\hskip 1em plus
	0.5em minus 0.4em\relax New York, {NY}, {USA}: Mc{G}raw-Hill, 2005.
	
	\bibitem{EGrossi2012}
	E.~{Grossi} and M.~{Lops}, ``{Space-time code design for MIMO detection based
		on Kullback-Leibler divergence},'' \emph{IEEE Transactions on Information
		Theory}, vol.~58, no.~6, pp. 3989--4004, 2012.
	
	\bibitem{SkolnikBook}
	M.~I. Skolnik, \emph{Introduction to radar systems}, 3rd~ed., ser. McGraw-Hill
	international editions. Electrical engineering series.\hskip 1em plus 0.5em
	minus 0.4em\relax Boston, MA, USA: McGraw-Hill, 2001.
	
	\bibitem{Bargallo-1994}
	J.~A. {Roberts} and J.~M. {Bargallo}, ``{DPSK performance for indoor wireless
		Rician fading channels},'' \emph{IEEE Transactions on Communications},
	vol.~42, no. 234, pp. 592--596, 1994.
	
	\bibitem{Reed-1999}
	S.~{Jonqyin} and I.~S. {Reed}, ``{Performance of MDPSK, MPSK, and noncoherent
		MFSK in wireless Rician fading channels},'' \emph{IEEE Transactions on
		Communications}, vol.~47, no.~6, pp. 813--816, Jun. 1999.
	
	\bibitem{Lange_2004}
	K.~Lange, ``A tutorial on {MM} algorithms,'' \emph{The American Statistician},
	vol.~58, no.~1, p. 30–37, Feb. 2004.
	
	\bibitem{Rennie_1991}
	B.~C. Rennie, ``Exponential means,'' \emph{James Cook Mathematical Notes},
	vol.~6, no.~54, pp. 6023--6024, Feb. 1991.
	
	\bibitem{Pauw_Schilling_1988}
	C.~K. {Pauw} and D.~L. {Schilling}, ``Probability of error of {$M$-ary PSK and
		DPSK on a Rayleigh} fading channel,'' \emph{IEEE Transactions on
		Communications}, vol.~36, no.~6, pp. 755--756, 1988.
	
	\bibitem{boyd2004convex}
	S.~Boyd, S.~P. Boyd, and L.~Vandenberghe, \emph{Convex optimization}.\hskip 1em
	plus 0.5em minus 0.4em\relax Cambridge, United Kingdom: Cambridge University
	Press, 2004.
	
	\bibitem{Gershman_2006}
	A.~B. Gershman, Z.-Q. Luo, and S.~Shahbazpanahi, ``Robust adaptive beamforming
	based on worst-case performance optimization,'' in \emph{Robust Adaptive
		Beamforming}, ian Li and P.~Stoica, Eds.\hskip 1em plus 0.5em minus
	0.4em\relax Hoboken, New Jersey: John Wiley \& Sons, 2006, ch.~2, pp. 49--89.
	
	\bibitem{1634819}
	N.~Sidiropoulos, T.~Davidson, and Z.-Q. Luo, ``Transmit beamforming for
	physical-layer multicasting,'' \emph{IEEE Transactions on Signal Processing},
	vol.~54, no.~6, pp. 2239--2251, 2006.
	
	\bibitem{Horn-2012}
	R.~A. Horn and C.~R. Johnson, \emph{Matrix Analysis}, 2nd~ed.\hskip 1em plus
	0.5em minus 0.4em\relax USA: Cambridge University Press, 2012.
	
\end{thebibliography}
\end{document}